\documentclass[preprint,onecolumn,showpacs,amsfonts,aps,prd,floatfix,nofootinbib,float,superscriptaddress]{revtex4} 
\usepackage{amsmath}
\usepackage{bm}
\usepackage{graphicx}
\usepackage{color}
\usepackage{booktabs}

\begin{document}
\title{A new scheme of causal viscous hydrodynamics for relativistic heavy-ion collisions: A Riemann solver for quark-gluon plasma}
\date{\today}

\author{Yukinao Akamatsu}
\affiliation{Kobayashi-Maskawa Institute for the Origin of Particles and the Universe (KMI), Nagoya University, Nagoya 464-8602, Japan}

\author{Shu-ichiro Inutsuka}
\affiliation{Department of Physics, Nagoya University, Nagoya 464-8602, Japan}

\author{Chiho Nonaka}
\affiliation{Kobayashi-Maskawa Institute for the Origin of Particles and the Universe (KMI), Nagoya University, Nagoya 464-8602, Japan}
\affiliation{Department of Physics, Nagoya University, Nagoya 464-8602, Japan}

\author{Makoto Takamoto}
\affiliation{Department of Physics, Nagoya University, Nagoya 464-8602, Japan}
\affiliation{Max-Planck-Institut f\"ur Kernphysik, Postfach 103980, 69029 Heidelberg, Germany}

\begin{abstract}
In this article, we present a state-of-the-art algorithm for solving the relativistic viscous 
hydrodynamics equation with the QCD equation of state. The numerical method is 
based on the second-order Godunov method and has less numerical dissipation, which 
is crucial in describing of quark-gluon plasma in high-energy heavy-ion collisions. 
We apply the algorithm to several numerical test problems such as sound wave propagation, shock tube and blast wave problems. 
In sound wave propagation, the intrinsic {\em numerical} viscosity is measured and 
its explicit expression is shown, which is the second-order of spatial resolution both 
in the presence and absence of {\em physical} viscosity. The expression of the numerical 
viscosity can be used to determine the maximum cell size in order to accurately measure 
the effect of physical viscosity in the numerical simulation.  
\end{abstract}

\pacs{}

\maketitle

\section{Introduction}
\label{sec:intro}
A relativistic fluid approach has been applied to various high-energy phenomena in astrophysics, nuclear, and hadron physics, bringing a lot of interesting and outstanding results. 
In particular, recent relativistic hydrodynamic analyses revealed a new and interesting feature of quark-gluon plasma (QGP) in high-energy heavy-ion collisions. 
Since the Relativistic Heavy-Ion Collider (RHIC) at Brookhaven National Laboratory (BNL) 
started operation in 2000, a number of discoveries have been made, providing insight into quantum chromodynamics (QCD) phase transition and the QGP.
One of the most interesting and surprising outcomes at RHIC was the production of the 
strongly interacting QGP (sQGP), which was confirmed by both theory and experiment. 
The highlights are: 
(i) strong elliptic flow, which suggests that collectivity and thermalization are achieved;
(ii) strong jet quenching, which confirms that hot and dense matter is created after collisions; (iii) the quark number scaling of elliptic flow, which indicates that the hot quark soup is produced \cite{white_papers, QGP}. 
Relativistic hydrodynamic models have made a significant contribution to these achievements.
For example, at the time, only hydrodynamic models could explain the strong elliptic flow at RHIC, which was considered to be direct evidence for the production of sQGP at RHIC. 
Because of the success of the relativistic hydrodynamic model at RHIC, hydrodynamic analysis has become a useful and powerful tool for understanding dynamics of hot and dense matter in high-energy heavy-ion collisions. 

In the early stage of the hydrodynamic studies at RHIC, viscosity effects were not taken into account.
However, detailed analyses of experimental data in relativistic heavy-ion collisions 
gradually revealed limitation of ideal hydrodynamic models. 
In Ref.~\cite{Romatschke:2007mq}, for the first time, quantitative analyses of elliptic flow were performed with a relativistic viscous hydrodynamic model.
The authors showed that ideal hydrodynamics overestimates elliptic flow as a function of transverse momentum, and that a hydrodynamic calculation with finite viscosity explains the experimental data better.
Since then, the main purpose of the phenomenological study for relativistic heavy-ion collisions at RHIC and LHC has been to obtain detailed information of bulk properties of QGP, such as its transport coefficients.
Besides, recent high statistical experimental data at RHIC and LHC require more rigorous 
numerical treatment on the hydrodynamical models. 
Recently both at RHIC and LHC the higher harmonic anisotropic flow, which is the Fourier coefficient of particle yield as a function of azimuthal angle, has been reported. 
One of the origins of the higher harmonics is event-by-event fluctuations. 
To obtain the precise value of transport coefficients with relativistic viscous hydrodynamics, 
we need to choose an algorithm with small numerical dissipation and treat the inviscid part with care. 
Usually each algorithm has advantages or disadvantages in terms of coding, computational time, precision and stability. 
Thus far, unfortunately, only limited attention has been paid to numerical aspects in hydrodynamic models for high-energy heavy-ion collisions.  

In this article, we present a state-of-the-art algorithm for solving the relativistic viscous hydrodynamics equation with the QCD equation of state (EoS). 
Our applications require a numerical scheme that can treat a shock wave appropriately and has less numerical dissipation in order to gain comprehensive understanding of recent high-energy heavy-ion collision physics.
These advantages can be achieved by implementing a Riemann solver for the relativistic ideal hydrodynamics.
In particular, we propose a new Riemann solver for the QCD EoS at low baryon density, which has not been considered in astrophysical application where baryon density is usually much higher.
We derive our Riemann solver by analytically solving the relativistic Riemann problem for low baryon density, within the approximation scheme proposed by \cite{Mignone:2005ns}.
As we will see in Section \ref{sec:num_test}, where we perform several numerical tests, our new algorithm with the Riemann solver has an advantage over other algorithms such as Kurganov-Tadmor (KT) \cite{Kurganov:2000},  Nessyahu-Tadmor (NT) \cite{Nessyahu:1990} and SHASTA \cite{SHASTA:1973} from the point of view of analyses for current relativistic heavy-ion collisions.
By implementing our new Riemann solver for relativistic ideal hydrodynamics in a numerical scheme for causal viscous hydrodynamics recently proposed in Ref.~\cite{Takamoto:2011wi}, we can also construct a new algorithm for causal viscous hydrodynamics for QGP.

This article is organized as follows.
In Section \ref{sec:hydro}, we review current hydrodynamic models for relativistic heavy-ion collisions and introduce the basics of relativistic hydrodynamics.
In Section \ref{sec:qcd_eos},
we explain the QCD EoS at high temperature and 
low baryon density based on the latest lattice QCD calculation. 
In Section \ref{sec:riemann}, we propose a new Riemann solver for the ideal fluid with the QCD EoS at high temperature and low baryon density.
In Section \ref{sec:num_test}, using the numerical scheme, 
we show results of several numerical tests, such as 
sound wave propagation, as well as shock tube and blast wave problems. 
Section \ref{sec:sum} is devoted to summary and discussions.
In this article, we adopt natural units, with the speed of light in vacuum $c=1$, Boltzmann constant $k_B=1$ and Planck's constant $\hbar=1$.

\section{Hydrodynamic models}
\label{sec:hydro}
First we list current hydrodynamic models, which are applied to relativistic heavy-ion collisions \cite{Nonaka:2012qw} in Tables~\ref{table:ideal-hydro} and \ref{table:viscous-hydro}.  
Here we mention the key aspects of numerical simulations in relativistic hydrodynamic models, which are classified into ideal versions and viscous ones. 
One of the important ingredients of hydrodynamic models is an EoS, needed for solving 
the relativistic hydrodynamics equation.
Different types of physics related to QCD phase transitions can be input into the EoS.
\footnote{
For the further application to relativistic heavy-ion collisions, not only EoS but also other aspects should be discussed; initial conditions and final conditions (freeze-out processes and final state interactions) of the hydrodynamic simulation.
Since modeling of these aspects is beyond the scope of this paper, they are not addressed in Tables~\ref{table:ideal-hydro} and \ref{table:viscous-hydro}.} 
From comparison between hydrodynamic calculations and experimental data of high-energy heavy-ion collisions, the information for the QCD phase diagram is obtained through the EoS used in the hydrodynamic calculation. 
The Bag model type EoS with the first-order phase transition has been widely used in relativistic hydrodynamic models, because of its simplicity and the lack of conclusive results on EoS of QCD.
In recent hydrodynamical calculations, lattice-inspired EoS has begun to be employed, thanks to the progress of thermodynamical analyses based on first principle calculations with lattice QCD simulation.

\begin{table}[htbp]
\caption{Ideal hydrodynamical models. In the table, we use the following abbreviation. 
lQCD: lattice QCD inspired EoS, SPH: smoothed particle hydrodynamics, 
PPM: piecewise parabolic method. 
}
\label{table:ideal-hydro}
\vspace{0.5cm}
\begin{tabular} {|c|c|c|c|}
\hline 
Ref. & Dimension & EoS & Numerical scheme \\
\hline \hline
Hama et al. \cite{Hama:2004rr} &  3+1 &   Bag model &  SPH    \\
\hline 
Hirano et al. \cite{Hirano:2005xf} & 3+1 & Bag model  & PPM \\ 
\hline 
Nonaka and Bass \cite{Nonaka:2006yn} & 3+1 &  Bag model & Lagrange    \\ 
\hline 
Hirano et al. \cite{Hirano:2010jg, Hirano:2010je} &  3+1&  lQCD & PPM \\ 
\hline 
Petersen et al. \cite{Petersen:2010cw} & 3+1&   lQCD  & SHASTA  \\
\hline 
Karpenko and Sinyukov \cite{Karpenko:2010te} & 3+1 & lQCD & HLLE \\ 
\hline 
Holopainen et al. \cite{Holopainen:2010gz} & 2+1 &  lQCD & SHASTA  \\
\hline 
Pang et al. \cite{Pang:2012he}  & 3+1 &  lQCD & SHASTA \\ 
\hline
\end{tabular}
\end{table}
\begin{table}[htbp]
\caption{Viscous hydrodynamical models. In the table, we use the following abbreviation. CD: central difference, and KT: Kurganov-Tadmor (KT) scheme.}
\label{table:viscous-hydro}
\vspace{0.5cm}
\begin{tabular} {|c|c|c|c|}
\hline 
Ref. & Dimension & EoS & Numerical scheme \\ \hline \hline
Romatschke and Romatschke \cite{Romatschke:2007mq} & 2+1 & lQCD  & CD    \\ \hline 
Luzum and Romatschke \cite{Luzum:2008cw} &  2+1     &      lQCD   &   CD         \\ \hline 
Schenke et al. \cite{Schenke:2010rr} & 3+1 &       lQCD         &    KT             \\ \hline 
Song et al. \cite{Song:2010mg} & 2+1 &  lQCD & SHASTA     \\ \hline
Chaudhuri \cite{Chaudhuri,Roy:2011pk}  & 2+1       & Bag model  & SHASTA             \\ \hline 
Bozek \cite{Bozek:2011ua} & 3+1 &      lQCD          &  CD             \\ \hline 
\end{tabular}
\end{table}

Another important ingredient in hydrodynamical models is a numerical scheme for 
solving the relativistic ideal and viscous hydrodynamical equations. 
Historically, in terms of analyses of high-energy heavy-ion collisions, only physical conditions, such as initial conditions, EoS and termination conditions of hydrodynamic expansion have been discussed. 
However, because of the nonlinearity of the relativistic hydrodynamics equations, even if we use the same physical conditions, 
different numerical schemes would give us different numerical solutions. 
Furthermore, when we start to investigate viscosity effects and event-by-event fluctuations in recent 
high statistic experimental data, we need to choose suitable numerical schemes carefully. 
For numerical stability of hydrodynamic calculation, numerical dissipation is needed. 
Therefore, in order to evaluate physical viscosity in high-energy heavy-ion collisions, 
we need to avoid or control the effect of numerical dissipation in the numerical relativistic viscous hydrodynamic calculation.
Accurate numerical schemes can be found in those with Riemann solvers for relativistic ideal hydrodynamics (references therein \cite{Takamoto:2011wi}). 
The Riemann solver is a method to calculate numerical flux by using the exact solution of the Riemann problems at the interfaces separating numerical grid cells, and can be used to describe the 
flows with strong shocks and sharp discontinuity stably and highly accurately.

Here, we mention the basis of hydrodynamics briefly.
The relativistic hydrodynamics equations are given by the conservation laws of energy, momentum and baryon number: 
\begin{eqnarray}
T^{\mu \nu}_{; \mu} (x)  & = & 0, 
\label{eq-hydro} \\
J_{B; \mu}^ {\mu } (x) & =& 0, 
\label{eq-net_baryon}
\end{eqnarray}
where $T^{\mu \nu}(x)$ is the energy-momentum tensor and $J_B^\mu(x)$ is the baryon current. 
Throughout this paper, we use the Cartesian coordinates where the metric tensor $g^{\mu \nu}$  is 
given by $g^{\mu \nu}= {\rm diag} (1, -1, -1, -1)$. 
In the case of relativistic ideal fluid, the energy-momentum tensor and baryon current are given by 
\begin{eqnarray}
T^{\mu \nu}(x)&=& [ e (x) + p(x) ] u^\mu (x) u^\nu  (x) - p(x) g^{\mu
    \nu} \, ,\\ 
J_B^\mu(x)&=&n_B (x)u^\mu(x)
\end{eqnarray}
where $e (x)$, $p(x)$, $n_B(x)$ and  $u^\mu (x)$ ($u^{\mu}(x)u_{\mu}(x)=1$) are the proper energy density,
pressure and  baryon density which are evaluated in the rest frame of the fluid and four-velocity, respectively. 

When the effects of dissipation are included into relativistic hydrodynamics,
a rather complicated situation arises. 
One of the difficulties is that the naive introduction of viscosities
as in the first-order theory, in which the entropy current contains no
terms higher than the first-order term in the thermodynamic fluxes,
suffers from acausality. 
In order to avoid this problem, the second-order terms in heat flow and 
viscosities have to be included in the expression for the entropy
\cite{Israel:1976tn,Muller,Dusling:2007gi,GO1,GO2,Ott,Muronga:2004sf}, but a systematic treatment of these second-order terms has not yet been established.
Although there has been remarkable progress toward the construction of a 
fully-consistent, relativistic viscous hydrodynamical theory for the description of high-energy heavy-ion collisions, there are still ongoing discussions about the formulation of 
the equations of motion and about the numerical procedures 
\cite{Teaney:2009qa}.

At first order the new structures are proportional to gradients
of the velocity field $u^\mu$ and the baryon number density $n_{\rm B}$, and only three proportionality constants appear:
the shear viscosity $\eta$, the bulk viscosity $\zeta$,
and the baryon number conductivity $\sigma$. 
At second order, many more new parameters related to relaxation phenomena, such as relaxation times for each diffusive modes $\tau_{\eta}$, $\tau_{\zeta}$, and $\tau_{\sigma}$ appear.
Currently, most viscous hydrodynamical calculations use the relativistic 
dissipative equations of motion that were derived phenomenologically 
by Israel and Stewart \cite{Israel:1976tn} which are utilized in this work (see Appendix \ref{sec:IS}), and their variants \cite{Muller,GO1,GO2,Ott, Dusling:2007gi,Muronga:2004sf}. 
Recently, a second-order viscous hydrodynamics from AdS/CFT correspondence was derived \cite{Baier:2007ix}, as well as a set of 
generalized Israel-Stewart equations from kinetic theory via Grad's 14-momentum 
expansion, which have several new terms \cite{Betz:2009zz}. 
However, a qualitatively different first-order relativistic dissipative hydrodynamical scheme was also proposed on the basis of renormalization-group consideration \cite{Tsumura:2007ji,Tsumura:2007wu}.

There are two choices for the local rest frame in a relativistic viscous hydrodynamics equation. 
One is the Eckart frame \cite{Eckart}, where the direction of the four-velocity is the same as that of the particle flux vector. 
The other is the Landau-Lifshitz frame \cite{Landau:fluid}, where the direction of the four-velocity is the same as that of energy flux vector. 
Because in high-energy collisions at RHIC and LHC the baryon number density is very small (Section \ref{sec:qcd_eos}), 
the Landau-Lifshitz frame is more suitable for QCD at high temperature and low baryon number density.

\section{QCD equation of state at high temperature and low baryon density}
\label{sec:qcd_eos}
\begin{figure}[t!]
\includegraphics[width=8cm,scale=1,angle=-90,clip]{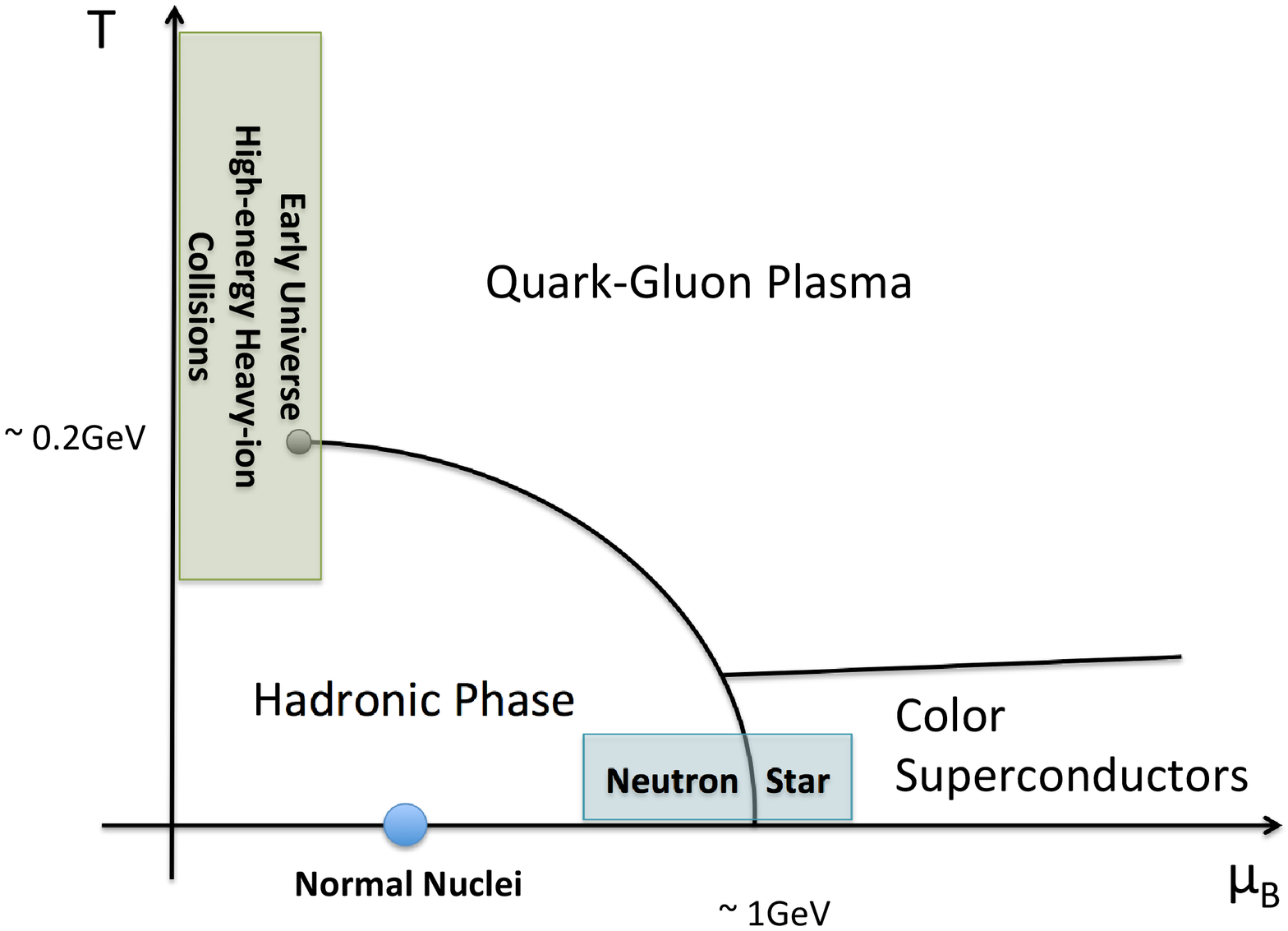}
\caption{
(Color online.)
A schematic QCD phase diagram.}
\label{fig:phase_diagram}
\end{figure}
The phase diagram of QCD matter has been investigated for decades.
In Fig. \ref{fig:phase_diagram}, a schematic QCD phase diagram is depicted with the axes of temperature $T$ and baryon chemical potential $\mu_{\rm B}$.
Among the six flavors of quarks in the Standard Model, we only consider the three light flavors of quarks (up, down, and strange) with physical quark masses.
The phase diagram is characterized by three typical phases: a hadronic phase, a quark-gluon plasma (QGP) phase, and a color super-conducting (CSC) phase.
In the hadronic phase, which is realized in the ground state of the QCD Hamiltonian (vacuum state), the chiral symmetry of QCD is broken, and quarks and gluons are confined in the hadrons.
In the QGP phase, which was realized in the early universe, the chiral symmetry is restored and quarks and gluons are liberated from the hadrons.
In the CSC phase, which may be realized inside neutron stars, the quarks on the Fermi surface form Cooper pairs and are condensed to create a super-conducting state.
For further details of the QCD phase diagram, see the review \cite{Fukushima:2010bq}.

In ultra-relativistic heavy-ion collisions at the LHC and RHIC, the relevant region in the QCD phase diagram is high-temperature ($T\sim 200 $-$1000$ MeV), low baryon density ($\mu_{\rm B}\sim 0 $-$100$MeV) one.
In this region, there is a transition from the QGP phase to the hadron phase.
The transition is a crossover confirmed by the state-of-the-art lattice QCD simulation \cite{Borsanyi:2010cj}, in contrast to the Bag EoS, which is a phenomenological equation of state with a first-order phase transition and has been widely utilized in previous hydrodynamic models.
In the high-temperature, low baryon density region, we expect that the QCD EoS can be approximated by taking into account the leading-order contribution of the finite baryon chemical potential.
In other words, due to the charge conjugation ($C$) symmetry of the QCD, the $C$-even quantities, e.g. pressure, energy density, temperature, and sound velocity, are approximated by those at vanishing baryon chemical potential, while the $C$-odd quantities, e.g. baryon density, and baryon chemical potential, are approximated by the first-order contribution of the chemical potential.
Note that in this approximation the $C$-even quantities are independent of $\mu_{\rm B}$, while the $C$-odd quantities depend on both $T$ and $\mu_{\rm B}$ in principle.
For example,
\begin{eqnarray}
\label{eq:eos_approx1}
p(T,\mu_{\rm B})&=&p(T,0) + \frac{1}{2}\chi(T,0)\mu_{\rm B}^2 + \mathcal O(\mu_{\rm B}^4) \approx p(T,0),\\
\label{eq:eos_approx2}
n_{\rm B}(T,\mu_{\rm B})&=&\frac{\partial p(T,\mu_{\rm B})}{\partial \mu_{\rm B}}=\chi(T,0)\mu_{\rm B} +  \mathcal O(\mu_{\rm B}^3)\approx \chi(T,0)\mu_{\rm B},
\end{eqnarray}
where $\chi$ stands for the baryon number susceptibility.
\footnote{
The approximated equation of state at high temperature and low baryon density satisfies the convexity condition for the relativistic hydrodynamics equations.
The fundamental derivative $\tilde{\mathcal G}$ is defined in terms of differentials of pressure along the isentropes:
\begin{eqnarray}
\tilde{\mathcal G}=-\frac{1}{2}\xi(1-c_{\rm s}^2)^2
\left(\frac{\partial^2 p}{\partial \xi^2}\right)_s \Bigg/\left(\frac{\partial p}{\partial \xi}\right)_s, \ \ \
\xi\equiv (e+p)/n_{\rm B}^2.
\end{eqnarray}
When $\tilde{\mathcal G}$ is positive, the convexity condition is satisfied \cite{Ibanez:2013bla}.
Our approximation corresponds to
$p(T,\mu_{\rm B})=\tilde p(s,\xi)=\tilde p_0(s) + \tilde p_1(s)\xi^{-1} + \cdots$ at $\xi\to\infty$,
and $\tilde{\mathcal G}\approx (1-c_{\rm s}^2)^2>0$ is easily confirmed at $\mu_{\rm B}\approx 0$.
}
Although the first-principles lattice QCD simulation is limited at vanishing baryon chemical potential, we can access the thermodynamic properties at low baryon chemical potential by using $\chi(T,0)$ in the above approximation.
Indeed, combining the result of the state-of-the-art lattice simulation \cite{Borsanyi:2010cj, Borsanyi:2011sw} ($p\approx 1$-$4 \ T^4$ at $T\approx200$-$1000$ MeV, $\chi\approx 0.2$-$0.3 \ T^2$ at $T\approx200$-$400$ MeV) and the typical values of the baryon chemical potential in the heavy-ion collisions ($\mu_{\rm B}\approx 24$ MeV at RHIC and 1 MeV at LHC \cite{Andronic:2011yq}), we can estimate the importance of the next-to-leading order term in the pressure by taking its ratio with the leading-order term $\chi(T,0)\mu_{\rm B}^2/p(T,0)\sim 0.3(\mu_{\rm B}/T)^2$, which yields only a 0.4\% correction at the RHIC and a 0.00075\% one at the LHC.
Therefore, we regard this approximation to be quantitatively reliable in all the regions of the QGP fireball at both RHIC and LHC.

\begin{figure}
\includegraphics[width=7cm,scale=1,angle=-90,clip]{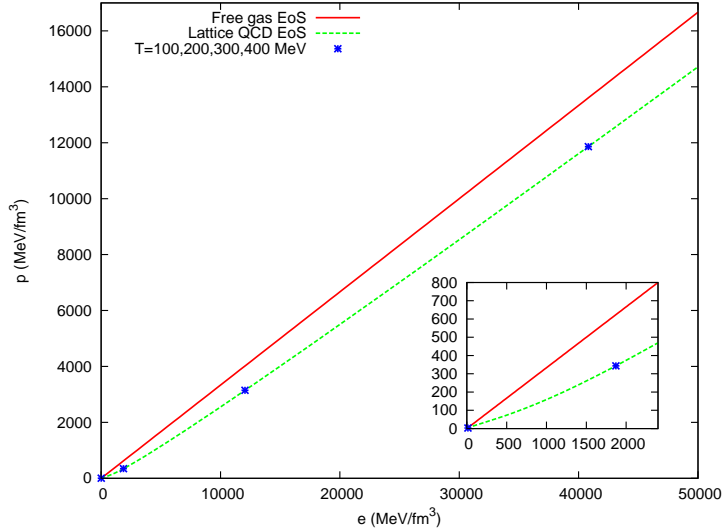}
\caption{
(Color online.)
Equation of states for free gas of gluons (free gas EoS) and interacting quarks and gluons calculated by the lattice QCD simulation (lattice QCD EoS) at vanishing baryon chemical potential.
For the latter, the energy densities and pressures at $T=100,200,300,$ and 400 MeV are plotted.
}
\label{fig:eos}
\end{figure}

In numerical tests in Section \ref{sec:num_test}, we will consider an EoS for free gas of gluons (free gas EoS) and that for realistic interacting quarks and gluons calculated by the lattice QCD simulation (lattice QCD EoS), which we plot in Fig.~\ref{fig:eos}.
In the free gas EoS, we adopt the parameterization of \cite{Molnar:2009tx}:
\begin{eqnarray}
e(T,0)=3p(T,0)=\frac{48T^4}{\pi^2},
\end{eqnarray}
to make comparison with other numerical schemes and introduce $\chi(T,0)=\epsilon T^2$ with $\epsilon\ll 1$ in order to achieve effectively gluonic matter without quarks.
In the lattice QCD EoS, we adopt the parameterization for the trace anomaly $I\equiv e-3p$ for (2+1) flavors given in Eq.~(3.1) and Table~2 of \cite{Borsanyi:2010cj}:
\begin{eqnarray}
p(T,0)&=&T^4\int^T_0\frac{dT'}{T'}\frac{I(T')}{T'^4},\\
\frac{I(T)}{T^4}&=&\exp(-h_1/t-h_2/t^2)\cdot\left(
h_0+\frac{f_0\cdot \left[\tanh(f_1\cdot t+f_2)+1\right]}{1+g_1\cdot t+g_2\cdot t^2}
\right),
\end{eqnarray}
with $t\equiv T/(200 \ {\rm MeV})$, $h_0=0.1396$, $h_1=-0.1800$, $h_2=0.0350$, $f_0=2.76$, $f_1=6.79$, $f_2=-5.29$, $g_1=-0.47$, and $g_2=1.04$.
We parameterize the baryon number susceptibility $\chi$ by fitting Fig.~7 and Table~1 of \cite{Borsanyi:2011sw}:
\begin{eqnarray}
\chi(T,0)&=& aT^2\left[1+\tanh\left(\frac{T-T_0}{\Delta T}\right)\right],\\
a &=& 0.15, \ T_0=167 \ {\rm MeV} , \ \Delta T = 60 \ {\rm MeV}.
\end{eqnarray}

\section{Riemann solver for ideal fluid}
\label{sec:riemann}
\subsection{Exact solution of the relativistic Riemann problem}
Riemann problem is a classic one-dimensional initial value problem in hydrodynamics with infinitesimal dissipation 
and plays an essential role in numerical hydrodynamics.
Since we are interested in QCD matter in extremely high temperatures, we restrict our discussion to the relativistic hydrodynamics \cite{Landau:fluid}.
The basic equations of the relativistic ideal hydrodynamics are the conservation equations for baryon number, momentum, and energy:
\begin{eqnarray}
\label{eq:conservation_laws}
\frac{\partial}{\partial t}
\left(\begin{array}{c}
D\\
\bm m\\
E
\end{array}\right)
+\bm{\nabla}\cdot
\left(\begin{array}{c}
D\bm{v}\\
\bm{m}\bm{v} + p\bm I\\
\bm{m}
\end{array}\right)
=0,
\end{eqnarray}
where $D,\bm{m},E$ are densities of baryon number, momentum, and energy; $p,\bm{v}$ are pressure and flow vector; and $\bm{I}$ is the identity matrix.
The relation between the conservative variables $\bm{U}\equiv (D,\bm{m},E)$ and primitive variables $\bm{V}\equiv (n_{\rm B},\bm{v},p)$ are
\begin{eqnarray}
\label{eq:pval_nB}
D&=&\gamma n_{\rm B},\\
\label{eq:pval_v}
\bm{m}&=&(e+p)\gamma^2\bm{v},\\
\label{eq:pval_p}
E&=&(e+p)\gamma^2-p,
\end{eqnarray}
where $\gamma\equiv(1-|\bm{v}|^2)^{-1/2}$ and $e=e(p,n_{\rm B})$ is given by the QCD EoS.

The initial condition of the Riemann problem is given by two uniform states separated by a discontinuity surface at $x=0$:
\begin{eqnarray}
\bm{V}(x,t=0)
=\Biggl\{\begin{array}{cc}
\bm V_L & (x<0) \\
\bm V_R & (x>0)
\end{array}.
\end{eqnarray}
The exact solution to this problem is constructed from three types of flows: shock wave, rarefaction wave, and contact discontinuity \cite{Pons:2000rm}.
In the solution, they evolve self-similarly and the wave structure depends only on $\xi\equiv x/t$ at $t>0$ (self-similar flow).
Shock wave is a discontinuous surface moving at a constant velocity $v_{\rm sh}$, across which physical states are related by Rankine-Hugoniot jump conditions:
\begin{eqnarray}
\label{eq:RH_cond1}
\left[1/D\right] &=&  -\zeta\left[v_x\right],\\
\label{eq:RH_cond2}
\left[(e+p)\gamma v_x\right] &=& \zeta\left[p\right],\\
\label{eq:RH_cond3}
\left[(e+p)\gamma v_{y,z}\right] &=& 0, \\
\label{eq:RH_cond4}
\left[(e+p)\gamma - p/D\right] &=& \zeta\left[pv_x\right],
\end{eqnarray}
where $[q]\equiv q-q_S$ denotes the difference between the preshock state $q_S$ $(S=L,R)$ and the postshock state $q$ and $\zeta\equiv\gamma_{\rm sh}/j$, $\gamma_{\rm sh}\equiv (1-v_{\rm sh}^2)^{-1/2}$, $j\equiv\gamma_{\rm sh}D_{S}(v_{\rm sh}-v_{x,S})$.
Strictly speaking, the physical structure of a shock wave can be described only by viscous hydrodynamics equations.
In the limit of infinitesimal viscosity, however, the structure of the shock wave becomes a discontinuous step function that can also be regarded as the weak solution of original differential equations for inviscid hydrodynamics.
Rarefaction wave is a continuous self-similar flow, through which physical states evolve by nonlinear ordinary differential equations:
\begin{eqnarray}
\label{eq:ODE_RW1}
(v_x-\xi)\frac{d n_{\rm B}}{d\xi}
+\{n_{\rm B}\gamma^2v_x(v_x-\xi)+n_{\rm B}\}\frac{dv_x}{d\xi} && \nonumber\\
+ \ n_{\rm B}\gamma^2v_y(v_x-\xi)\frac{dv_y}{d\xi}
+n_{\rm B}\gamma^2v_z(v_x-\xi)\frac{dv_z}{d\xi}&=&0,\\
\label{eq:ODE_RW2}
(e+p)\gamma^2(v_x-\xi)\frac{dv_x}{d\xi}+(1-v_x\xi)\frac{dp}{d\xi}&=&0,\\
\label{eq:ODE_RW3}
(e+p)\gamma^2(v_x-\xi)\frac{dv_{y,z}}{d\xi}+v_{y,z}\xi\frac{dp}{d\xi}&=&0.
\end{eqnarray}
Contact discontinuity is also a discontinuous surface, across which  pressure $p$ and the flow velocity $v_x$ are continuous while other variables are  discontinuous in general.

\begin{figure}[t!]
\includegraphics[width=8cm,scale=0.5,clip]{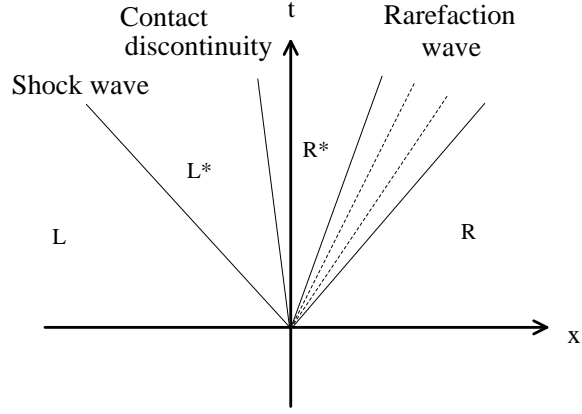}
\caption{
An example of the exact solution for a Riemann problem.
The regions $L$, $L^*$, $R^*$, and $R$ are uniform and separated by a shock wave, a contact discontinuity, and a rarefaction wave, respectively.
The pressures in the emergent intermediate states $L^*$ and $R^*$ are the same.
The region of the rarefaction wave is depicted by dotted lines.
}
\label{fig:Riemann}
\end{figure}

The exact solution to the relativistic Riemann problem with general initial condition is given by Pons {\it et al.} \cite{Pons:2000rm}, which we summarize as follows.
See also Fig.~\ref{fig:Riemann} as an example.
\begin{enumerate}
\item Connect the initial left uniform state ($L$) and emergent uniform state inside ($L^*$) by a shock wave or a rarefaction wave that propagates toward $L$.
\item Connect the initial right uniform state ($R$) and emergent uniform state inside ($R^*$) by a shock wave or a rarefaction wave that propagates toward $R$.
\item Connect the emergent uniform states ($L^*, R^*$) by a contact discontinuity.
\end{enumerate}
For the third step to hold, the emergent uniform states must be chosen so that the pressure $p$ and the flow velocity $v_x$ is continuous across the contact discontinuity.
Whether $L$ and $L^*$ ($R$ and $R^*$) are connected by a shock wave or a rarefaction wave depends on the the initial condition.
In general, when the pressure of the emergent intermediate state is higher (lower) than the initial pressure of each side, the shock (rarefaction) wave propagates toward that side.
Actual pressure of the emergent intermediate state can be known only after solving the Riemann problem.

\subsection{Approximation scheme for the low-density QCD equation of state}
In numerical scheme with the exact Riemann solver, we have to solve all the independent Riemann problems defined at the boundaries of all the two adjacent cells, but numerical solution of the ordinary differential equation for the rarefaction wave Eqs.\eqref{eq:ODE_RW1}-\eqref{eq:ODE_RW3} costs a lot of computational time.
Mignone {\it et al.} \cite{Mignone:2005ns} proposed an efficient approximation scheme to the exact solution of the Riemann problem.
In their scheme, the rarefaction waves are approximated by the discontinuity that satisfies conservation laws.
However, the original approximation scheme \cite{Mignone:2005ns} needs to be modified for the QCD matter with low baryon density because it frequently uses the specific enthalpy $h\equiv (e+p)/n_{\rm B}$ which diverges in vanishing baryon density $n_{\rm B}=0$.
Moreover, the their model EoSs, whose analytical simplicity also accelerates the numerical calculation, do not directly fit to the QCD EoS in low baryon density.

Here we present a new approximation scheme for the QCD matter with low baryon density.
First the Rankine-Hugoniot jump conditions \eqref{eq:RH_cond1}-\eqref{eq:RH_cond4} yield the Taub adiabat \cite{Taub:1948zz}
\begin{eqnarray}
\left[\left(\frac{e+p}{n_{\rm B}}\right)^2\right]
=\left(\frac{e(p;S)+p}{n_{\rm B}^2(p;S)}+\frac{e_S+p_S}{n_{{\rm B},S}^2}\right)[p],
\end{eqnarray}
where $\bm V(p;S)$ $(S=L,R)$ denotes the postshock variables, that is, the variables in the emergent uniform states $(L^*,R^*)$.
Once we specify a trial postshock pressure $p$ and approximate the QCD EoS by $e=e(p,n_{\rm B})\approx e(p,n_{\rm B}=0)$ as given in Section \ref{sec:qcd_eos}
\footnote{
If the baryon density is not low, one must solve $e(p;S)=e(p,n_{\rm B}(p;S))$ together with Eq.~\eqref{eq:Taub_nB} to obtain $e(p;S)$ and $n_{\rm B}(p;S)$, which requires 
additional iteration for solving this equation.
}, 
we can solve the Taub adiabat for the postshock variables $\bm V(p;S)$ with the preshock variables $\bm V_S$, that is, the variables in the initial uniform states $(L,R)$:
\begin{eqnarray}
e(p;S)&=&e(p,n_{\rm B}=0),\\
n_{\rm B}(p;S)&=&n_{{\rm B},S}\sqrt{\frac{\left\{e(p;S)+p_S\right\}\left\{e(p;S)+p\right\}}{(e_S+p_S)(e_S+p)}}.
\label{eq:Taub_nB}
\end{eqnarray}
The baryon flux across the shock $j(p;S)$ is also solved:
\begin{eqnarray}
j^2(p;S)&=&-\frac{\left[p\right]}{\left[(e+p)/n_{\rm B}^2\right]}\nonumber \\
&=&n_{{\rm B},S}^2\frac{e(p;S)+p_S}{e_S+p_S}\frac{[p]}{[e]-[p]}.
\end{eqnarray}
For later convenience, we define normalized baryon flux $J(p;S)\equiv j(p;S)/n_{{\rm B},S}$:
\begin{eqnarray}
J^2(p;S)=\frac{e(p;S)+p_S}{e_S+p_S}\frac{[p]}{[e]-[p]},
\end{eqnarray}
with which the flow velocity $v_x(p;S)$ is given by
\begin{eqnarray}
v_x(p;S)&=&\frac{(e_S+p_S)\gamma_S^2v_{x,S} + [p]\zeta(p;S)}
{(e_S+p_S)\gamma_S^2 + [p]\left\{v_{x,S}\zeta(p;S)+1 \right\}},\\
\zeta(p;S)&=&\frac{v_{x,S}\pm \sqrt{1+(1-v_{x,S}^2)\gamma_S^2/J^2(p;S)}}{1-v_{x,S}^2}.
\end{eqnarray}
In practice, the limit $[p]\to 0$ and $[e]\to 0$ in the normalized baryon flux $J(p;S)$ becomes numerically inaccurate.
Therefore, when $[p]$ or $[e]$ is tiny, we switch to the analytical limiting value:
\begin{eqnarray}
\lim_{[p],[e]\to0}J^2(p;S)
=\frac{c_{\rm s}^2(p;S)}{1-c_{\rm s}^2(p;S)}
=\frac{c_{{\rm s},S}^2}{1-c_{{\rm s},S}^2},
\end{eqnarray}
where $c_{\rm s}(p;S)=c_{\rm s}(p,n_{\rm B}=0)$ is the sound velocity.
The sign in $\zeta(p;S)$ is chosen to be $+(-)$ for $S=R(L)$ so that the correct shock propagation is ensured when the approximation scheme gives an exact solution.

Since the two postshock states are separated by a contact discontinuity, pressure $p$ and the flow velocity $v_x$ must be continuous.
Therefore we have to solve
\begin{eqnarray}
\label{eq:contact_disc}
v_x(p;L)=v_x(p;R),
\end{eqnarray}
whose solution $p^*$ gives $v^*_x\equiv v_x(p^*;L)=v_x(p^*;R)$ and other postshock variables.
This part is solved numerically by the Newton-Raphson algorithm with the following iteration:
\begin{eqnarray}
p^{(n+1)}&=&p^{(n)}-\frac{v_x(p^{(n)};L)-v_x(p^{(n)};R)}{v'_x(p^{(n)};L)-v'_x(p^{(n)};R)},\\
v'_x(p;S)&\equiv& \frac{dv_x(p;S)}{dp}\nonumber\\
&=&\frac{\left\{\zeta(p;S)+[p]\zeta'(p;S)\right\}\left\{1-v_{x,S}v_x(p;S)\right\}-v_x(p;S)}
{(e_S+p_S)\gamma_S^2 + [p]\left\{v_{x,S}\zeta(p;S)+1\right\}},\\
\left[p\right]\zeta'(p;S)&=&-\frac{1}{2}\gamma_S^2
\frac{n_{{\rm B},S}^2\frac{d}{dp}\left(\frac{e+p}{n_{\rm B}^2}\right)_{\rm Taub}+\frac{1}{J^2(p;S)}}{\zeta(p;S)(1-v_{x,S}^2)-v_{x,S}},\\
n_{{\rm B},S}^2\frac{d}{dp}\left(\frac{e+p}{n_{\rm B}^2}\right)_{\rm Taub}
&\equiv&n_{{\rm B},S}^2\frac{d}{dp}\left(\frac{e(p;S)+p}{n_{\rm B}^2(p;S)}\right)\nonumber\\
&=&\frac{e_S+p_S}{e(p;S)+p_S}\left(1-\frac{e_S+p}{e(p;S)+p_S}\frac{1}{c_{\rm s}^2(p;S)}\right).
\end{eqnarray}
This is the new approximation scheme for QCD matter with low baryon density.
It is evident that there is no singularity in the limit $n_{\rm B}\to 0$ in the new approximation scheme.

\subsection{Primitive recovery}
Once the solution $p^*$ for Eq.~\eqref{eq:contact_disc} is obtained, the numerical flux is determined and the system is evolved according to the relativistic ideal hydrodynamics equation \eqref{eq:conservation_laws}.
Since the time evolution by Eq.~\eqref{eq:conservation_laws} updates the conserved variables $\bm U=(D,\bm m, E)$ at each time step, we need to find a solution for the primitive variables $\bm V=(n_{\rm B},\bm v,p)$ by Eqs.~\eqref{eq:pval_nB}, \eqref{eq:pval_v}, and \eqref{eq:pval_p} with the given updated $\bm U$.
The problem is reduced to solving
\begin{eqnarray}
f(p)&\equiv& \left[e(p,n_{\rm B}(p)) 
+ p\right]\gamma^2(p)-E-p=0, \\
\frac{1}{\gamma^2(p)} &\equiv& 1-\frac{\bm m^2}{(E+p)^2},\ \ 
n_{\rm B}(p) \equiv \frac{D}{\gamma(p)}.
\end{eqnarray}
The Newton-Raphson algorithm for numerically solving $f(p)=0$ is given by the following iteration:
\begin{eqnarray}
p^{(n+1)}&=&p^{(n)}-\frac{f(p)}{df(p)/dp}, \\
\frac{df(p)}{dp}
&=&\left\{\frac{\partial e(p,n_{\rm B})}{\partial p}+1\right\}
\gamma^2(p)-1 \nonumber \\
&& +\left\{
\frac{\partial e(p,n_{\rm B})}{\partial n_{\rm B}}\frac{D}{\gamma(p)}
-2\left(e(p,n_{\rm B})+p\right)
\right\}
\frac{\gamma^2(p)}{E+p}\left(\gamma^2(p)-1\right).
\end{eqnarray}
In the low baryon density region, we get the following expression for the partial derivatives of $e(p,n_{\rm B})$:
\begin{eqnarray}
\frac{\partial e(p,n_{\rm B})}{\partial p}
&=&\frac{1}{c^2_{\rm s}(p,n_{\rm B}=0)}+\mathcal O(n^2_{\rm B})
\approx \frac{1}{c^2_{\rm s}(p,0)},\\
\frac{\partial e(p,n_{\rm B})}{\partial n_{\rm B}}
&=&\frac{n_{\rm B}}{\chi(T,\mu_{\rm B}=0)}\left(1+\frac{T}{\chi(T,0)}\frac{\partial \chi(T,0)}{\partial T}-\frac{1}{c^2_{\rm s}(p,0)}\right)
+\mathcal O(n_{\rm B}^3)\nonumber \\
&\approx& \frac{n_{\rm B}}{\chi(T,0)}\left(1+\frac{T}{\chi(T,0)}\frac{\partial \chi(T,0)}{\partial T}-\frac{1}{c^2_{\rm s}(p,0)}\right).
\end{eqnarray}
In this region, it is sufficient to solve $p(T,\mu_{\rm B}=0)=p$ to get the temperature, which is needed to calculate $\partial \chi(T,0)/\partial T$.
\footnote{
In this section, we express the thermodynamic quantities as functions of $(p,n_{\rm B})$.
However $\partial \chi(T,0)/\partial T$ is not {\it conveniently} expressed by such functions, we here write the susceptibility $\chi$ as a function of ($T,\mu_{\rm B}$).
}
By this algorithm for primitive recovery, the relativistic ideal hydrodynamics can also be solved.

\section{Numerical Tests}
\label{sec:num_test}
By applying the Riemann solver in Section \ref{sec:riemann} to the numerical scheme of causal viscous hydrodynamics \cite{Takamoto:2011wi}, we solve several test problems, namely sound wave propagation, shock tube and blast wave problems in both ideal and viscous hydrodynamics.
The structure of numerical algorithm is reviewed in Appendix \ref{sec:numerical_algorithm}.

\subsection{Sound wave propagation}
\subsubsection{L1 norm as a measure of accuracy}

\begin{figure}[t!]
\includegraphics[width=5cm,scale=0.5,angle=-90,clip]{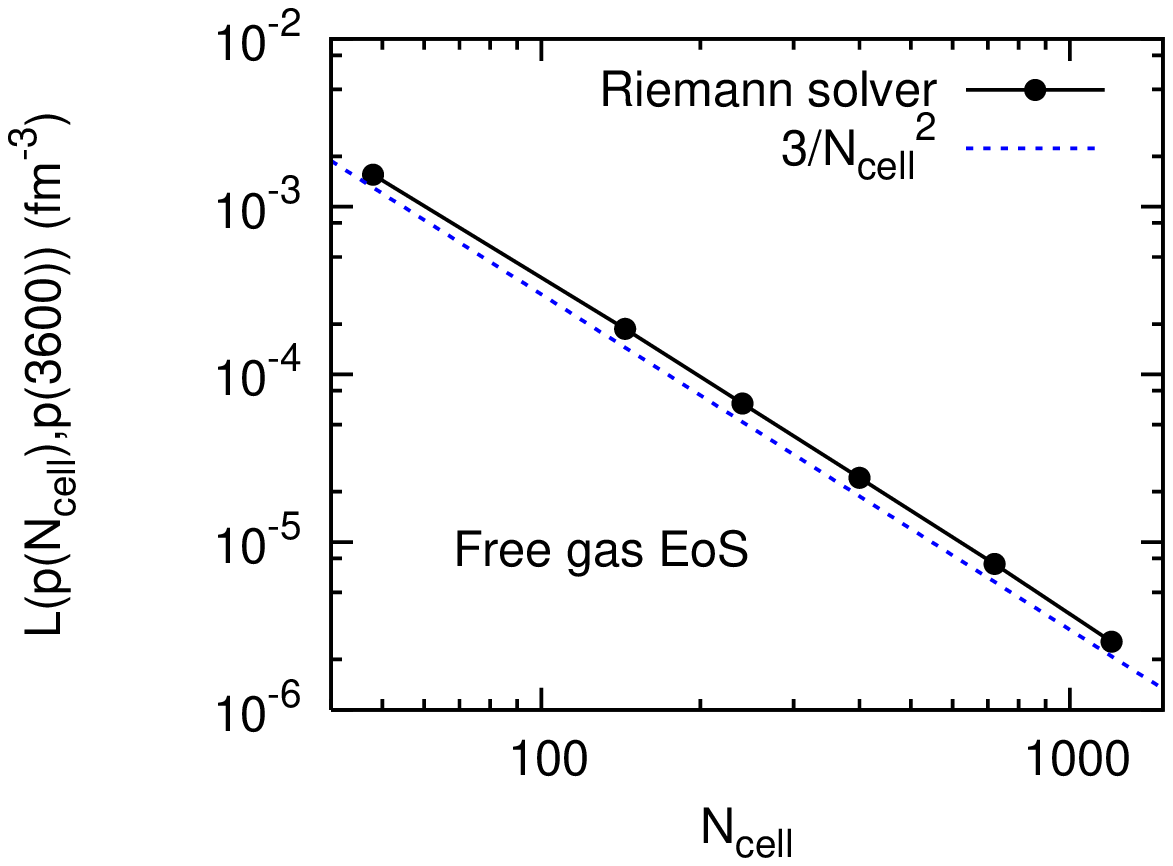}
\includegraphics[width=5cm,scale=0.5,angle=-90,clip]{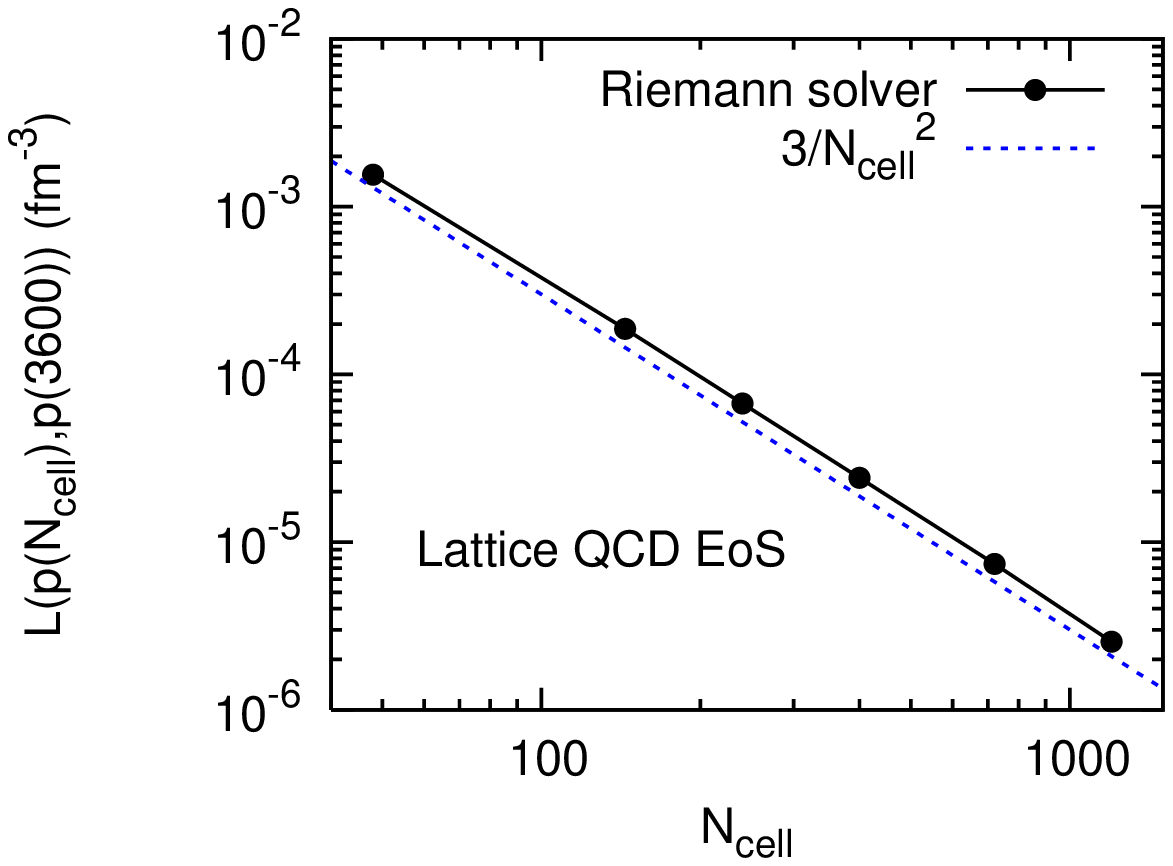}
\caption{
(Color online.)
L1 norm for pressure at $t=\lambda/c_{\rm s0}$ for the free gas EoS (left) and for the lattice QCD EoS (right).
The dotted line indicates a scaling $L(p(N_{\rm cell}),p(3600))\propto 1/N_{\rm cell}^2$.
}
\label{fig:L1norm}
\end{figure}

Here we perform a simulation of sound wave propagation in ideal hydrodynamics using the numerical scheme presented above.
The system length is $L_x=\lambda=2$ fm and is discretized with $N_{\rm cell}=48, 144, 240, 400, 720, 1200$, and 3600 cells.
We set an initial condition 
\begin{eqnarray}
\label{eq:init}
{\bm V}(x,t=0)
= \left(
0,
\frac{\delta p}{c_{\rm s0}(e_0+p_0)}\sin\left(2\pi x/\lambda\right),
0,0,
p_0+\delta p\sin\left(2\pi x/\lambda\right)
\right)
\equiv {\bm V_{\rm init}}(x),
\end{eqnarray}
and impose a periodic boundary condition $\bm V(-\lambda/2,t)=\bm V(\lambda/2,t)$.
Here $e_0\equiv e(p_0,n_{\rm B}=0)$, $c_{\rm s0}\equiv c_{\rm s}(p_0,n_{\rm B}=0)$ and $p_0=10^3 \ {\rm fm^{-4}}$ and $\delta p=10^{-1} \ {\rm fm^{-4}}$.
Since the amplitude of the wave is small $\delta p/p_0=10^{-4} \ll 1$, the nonlinear hydrodynamics equation is approximated by linearized hydrodynamics equation, which possesses a sound wave mode $\bm V_{\rm s}(x,t)=(n_{{\rm B}\rm s}(x,t),\bm{v}_{\rm s}(x,t),p_{\rm s}(x,t))=\bm V_{\rm init}(x-c_{\rm s0} t)$ as its solution.
\footnote{
The accuracy of linear approximation can be discussed as follows.
In a rough estimate, the nonlinearity of ideal hydrodynamics equation is parameterized by $\epsilon\sim (\delta p/p_0)\cdot t/(\lambda/c_{\rm s0})=\mathcal O(10^{-4})$ and the sound wave solution is different from the exact solution by $\delta p\cdot\epsilon$.
Therefore as far as the L1 norm equation \eqref{eq:L1norm1} is larger than $\delta L\sim\delta p\cdot \epsilon\cdot\lambda\sim {\mathcal O(10^{-5})} \ [{\rm fm^{-3}}]$, the linearized sound wave solution can be practically regarded as the exact solution, which is the case for $N_{\rm cell}< 500$ in Fig.~\ref{fig:L1norm}.
}
We analyze the precision of our numerical scheme and its dependence on $N_{\rm cell}(<3600)$ by calculating the L1 norm for pressure after one cycle $t=\lambda/c_{\rm s0}$:
\begin{eqnarray}
\label{eq:L1norm1}
L(p(N_{\rm cell}),p(3600))=
\sum_{i=1}^{N_{\rm cell}}\mid p(x_i,\lambda/c_{\rm s0};N_{\rm cell})
-p(x_i,\lambda/c_{\rm s0};3600)\mid\frac{\lambda}{N_{\rm cell}}.
\end{eqnarray}
We expect a scaling $L(p(N_{\rm cell}),p(3600))\propto (\delta p/N^2_{\rm cell})\cdot N_{\rm cell}\cdot (\lambda/N_{\rm cell})=\lambda\delta p/N_{\rm cell}^2$ after one cycle since our numerical scheme is of second-order accuracy with respect to space and time discretization.
\footnote{
The precision after one cycle $\delta p/N_{\rm cell}^2$ is independent of the wavelength $\lambda$ and the sound velocity $c_{\rm s0}$.
As far as the linear approximation to the original full hydrodynamics equation works, any sound wave problem is identical to a single problem by scaling $t=(\lambda/c_{\rm s0} )t'$ and $x=\lambda x'$.
Since $N_{\rm cell}$ is fixed and so is the number of time steps after one cycle with the same Courant number ($c_{\rm s}\Delta t/\Delta x=0.1$ in this analysis), the precision is independent of $\lambda$ and $c_{\rm s0}$.
}
The results of the L1 norm for the free gas EoS and the lattice QCD EoS are shown in Fig.~\ref{fig:L1norm}.
We indeed find a scaling $L(p(N_{\rm cell}),p(3600))\propto 1/N^2_{\rm cell}$ for both equations of states, which is consistent with the theoretical expectation.

\begin{figure}[t!]
\includegraphics[width=6cm,scale=0.5,angle=-90,clip]{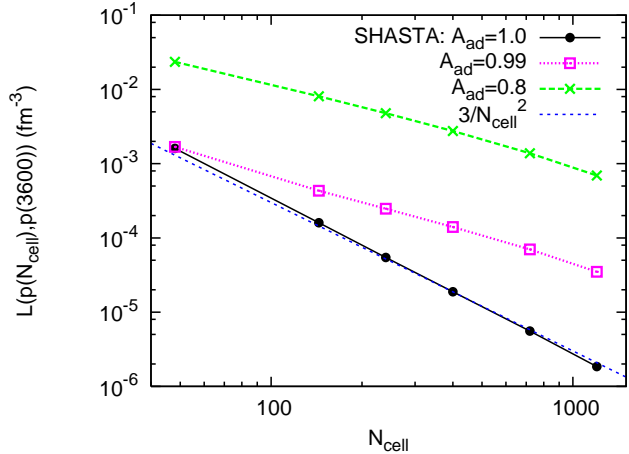}\caption{
(Color online.)
L1 norm for pressure at $t=\lambda/c_{\rm s0}$ for the free gas EoS calculated by SHASTA scheme with $A_{\rm ad}=1.0$, 0.99, and 0.8.
The dotted line indicates a scaling $L(p(N_{\rm cell}),p(3600))\propto 1/N_{\rm cell}^2$ for $A_{\rm ad}=1.0$.
We can also see $L(p(N_{\rm cell}),p(3600))\propto 1/N_{\rm cell}$ for $A_{\rm ad}=0.99$ and 0.8.
}
\label{fig:L1norm_shasta}
\end{figure}

We repeat the same analyses of the sound wave propagation using SHASTA algorithm \cite{SHASTA:1973} for the relativistic ideal hydrodynamics.
In this calculation, we only adopted the free gas EoS.
The result of the L1 norm is shown in Fig.~\ref{fig:L1norm_shasta}.
We find that the numerical accuracy is quite sensitive to the choice of the anti-diffusion parameter $A_{\rm ad}$ in the code.
With the anti-diffusion parameter $A_{\rm ad}=1.0$, we find that the SHASTA scheme not only exhibits the second-order accuracy but also has quantitatively similar accuracy to the algorithm based on our Riemann solver.
On the other hand, with $A_{\rm ad}=0.99$ and 0.8, the SHASTA scheme only exhibits the first-order accuracy and the L1 norm is quite large compared to that with $A_{\rm ad}=1.0$ with the same grid size.
The anti-diffusion parameter $A_{\rm ad}$ is introduced to reduce the numerical dissipation.
The default value $A_{\rm ad}=1.0$ minimizes the numerical dissipation due to the finite cell size when the system is smooth.
However, numerical accuracy of a scheme must be discussed together with its stability required by a problem to be solved.
This will be discussed in the next numerical test of shock tube problem.

\subsubsection{Numerical dissipation}
The simulation of sound wave propagation can also be utilized to estimate the numerical dissipation of the scheme.
Since any numerical scheme introduces tiny numerical dissipation, the sound wave in the simulation is attenuated even without physical viscosity.
The value of numerical dissipation is evaluated by the value of physical shear viscosity which gives the same amount of sound wave attenuation in the linearized region.
By linear analysis, the dispersion relation of the sound mode in viscous hydrodynamics with $n_{\rm B}=0, \ \zeta=\sigma=0$ is \cite{Baier:2007ix}
\begin{eqnarray}
\label{eq:dispersion}
\omega &=& \pm c_{\rm s0} k - i\gamma k^2 + {\mathcal O}(k^3),\\
\gamma &\equiv& \frac{2\eta}{3(e_0+p_0)}.
\end{eqnarray}
Note that the dispersion relation is independent of the relaxation time for shear mode $\tau_{\eta}$ in long wavelength limit.
The amplitude of sound wave with wave length $\lambda=2\pi/k$ is decreased by a factor of $\exp\left[-\frac{8\pi^2\eta}{3\lambda c_{\rm s0}(e_0+p_0)}\right]$ after one cycle ($t=\lambda/c_{\rm s0}$):
\begin{eqnarray}
p_{\rm s}(x,\lambda/c_{\rm s0};\eta)-p_0
=\left[p_{\rm s}(x,\lambda/c_{\rm s0})-p_0\right]
e^{-\frac{8\pi^2\eta}{3\lambda c_{\rm s0}(e_0+p_0)}}.
\end{eqnarray}
To quantify the attenuation of the sound wave due to the viscosity, let us utilize the L1 norm and define the numerical dissipation $\eta_{\rm num}$:
\begin{eqnarray}
\label{eq:L1norm_lin}
L(p_{\rm s}(\eta),p_{\rm s})
&=&\int_{-\lambda/2}^{\lambda/2} dx
\mid p_{\rm s}(x,\lambda/c_{\rm s0};\eta)
- p_{\rm s}(x,\lambda/c_{\rm s0})\mid \nonumber \\
&=&\frac{2\lambda\delta p}{\pi}
\left[1-e^{-\frac{8\pi^2\eta}{3\lambda c_{\rm s0}(e_0+p_0)}}\right] \equiv L_{\rm lin}(\eta),\\
L(p(N_{\rm cell}),p_{\rm s})&=&
\sum_{i=1}^{N_{\rm cell}}\mid p(x_i,\lambda/c_{\rm s0};N_{\rm cell})
-p_{\rm s}(x_i,\lambda/c_{\rm s0})\mid\frac{\lambda}{N_{\rm cell}},\\
L_{\rm lin}(\eta_{\rm num})&\equiv&L(p(N_{\rm cell}),p_{\rm s}),
\end{eqnarray}
by which we obtain
\begin{eqnarray}
\label{eq:num_vis}
\eta_{\rm num}= -\frac{3\lambda}{8\pi^2}c_{\rm s0}(e_0+p_0)
\ln\left[1-\frac{\pi}{2\lambda\delta p}L(p(N_{\rm cell}),p_{\rm s})\right].
\end{eqnarray}

\begin{figure}[t!]
\includegraphics[width=5cm,scale=0.5,angle=-90,clip]{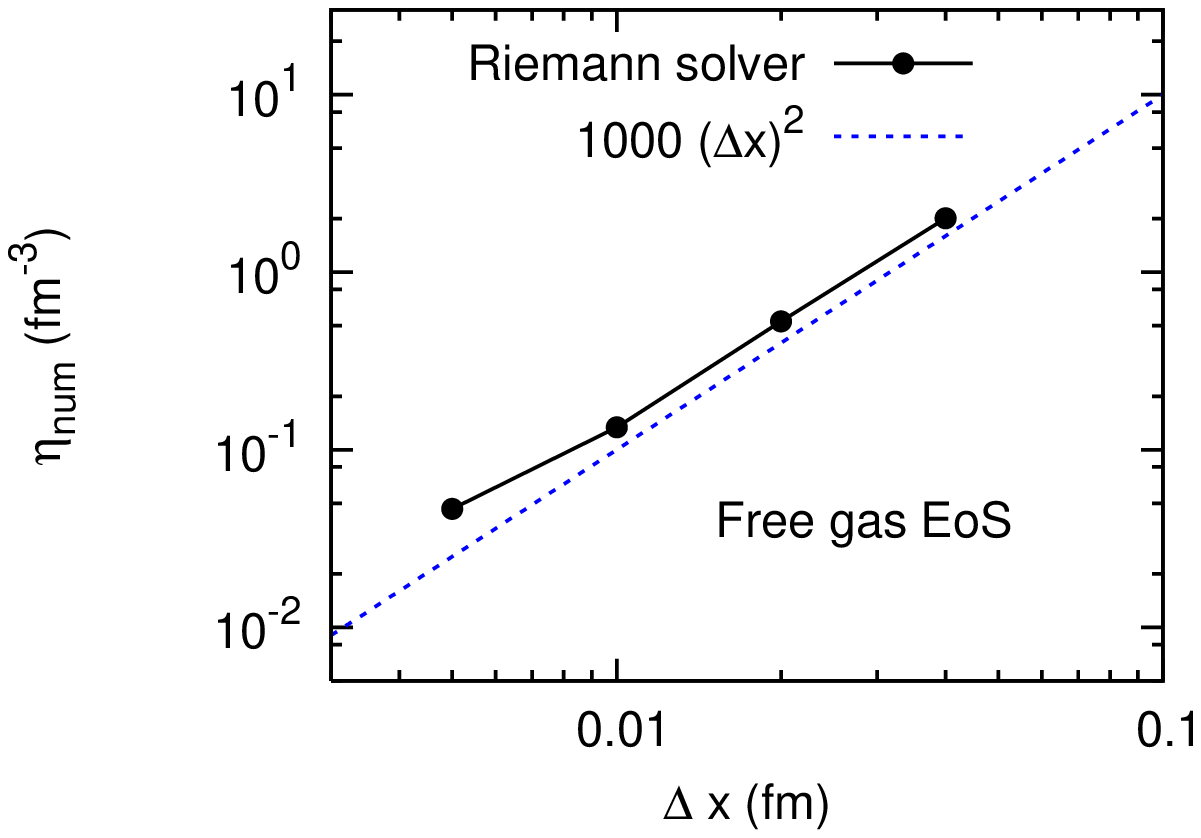}
\includegraphics[width=5cm,scale=0.5,angle=-90,clip]{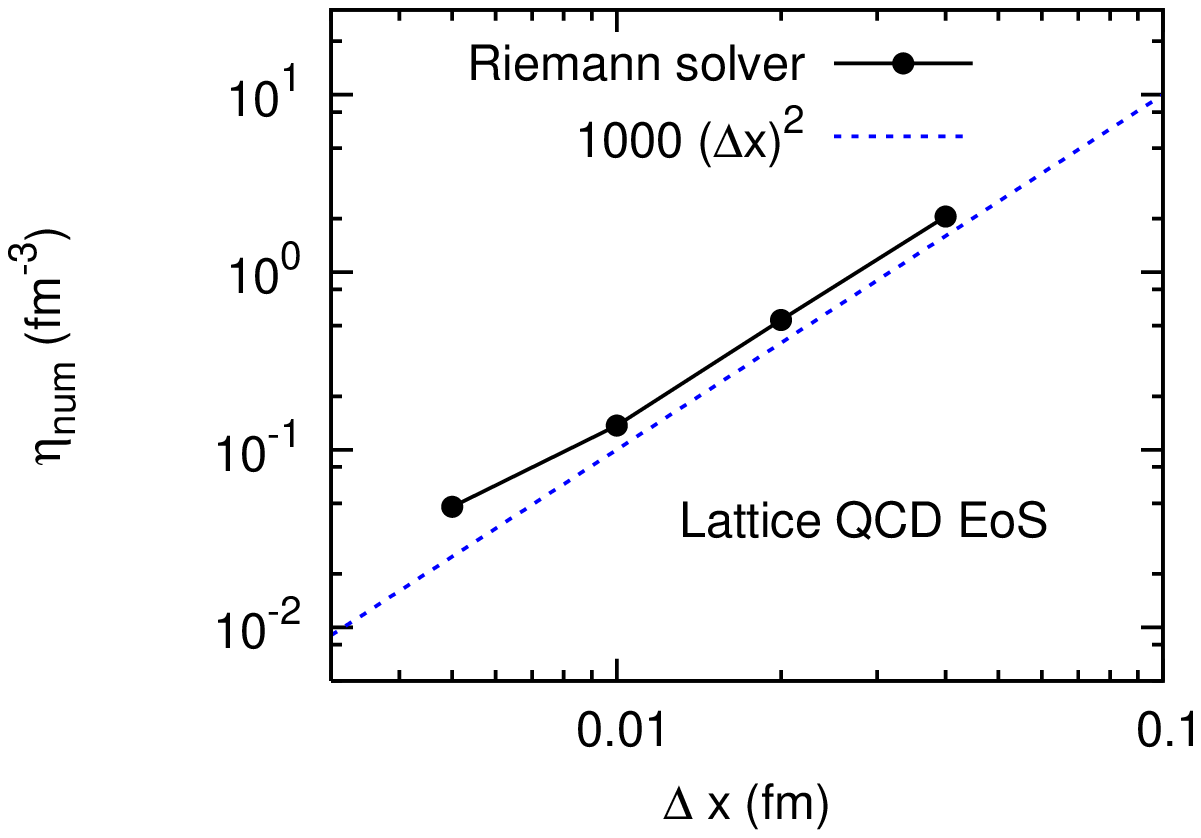}
\caption{
(Color online.)
Numerical dissipation as a function of cell size $\Delta x$ for (left) the free gas EoS and (right) the lattice QCD EoS.
The dotted line indicates $\eta_{\rm num}\approx 1000(\Delta x)^2$ for both EoSs.
}
\label{fig:num_vis}
\end{figure}

In Fig.~\ref{fig:num_vis}, we show the numerical dissipation of our scheme for the free gas EoS and for the lattice QCD EoS.
In these calculations, we choose $N_{\rm cell}=50,100,200,$ and 400, for which the linearized sound wave solution is precise enough as an approximation to the exact solution of the nonlinear hydrodynamics equations.
For both EoSs, the numerical dissipation can be approximated by $\eta_{\rm num}\approx 1000(\Delta x)^2$, where $\Delta x=\lambda/N_{\rm cell}$.
From Eq.~\eqref{eq:num_vis} and the second-order accuracy of our numerical scheme $L(p(N_{\rm cell}),p_{\rm s})\propto \lambda\delta p/N_{\rm cell}^2=(\delta p/\lambda)\cdot (\Delta x)^2$, the numerical dissipation is expected to scale with $\eta_{\rm num}\propto \left[c_{\rm s0}(e_0+p_0)/\lambda\right]\cdot (\Delta x)^2$.
Using the values of $p_0$, $e_0$, $c_{\rm s0}$, and $\lambda$, we find
\begin{eqnarray}
\eta_{\rm num}\approx 1\cdot\frac{c_{\rm s0}(e_0+p_0)}{\lambda} (\Delta x)^2
\end{eqnarray}
for both EoSs.

\begin{figure}[t!]
\includegraphics[width=5cm,scale=0.5,angle=-90,clip]{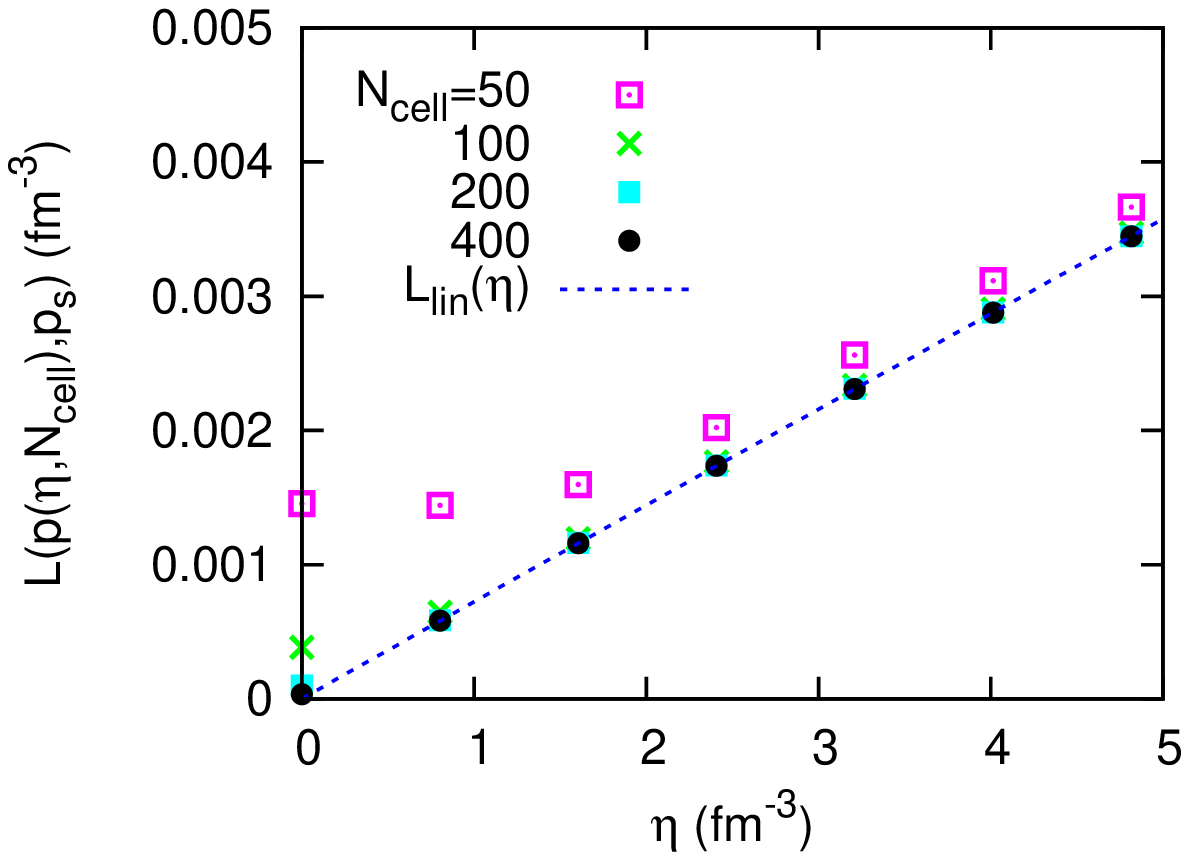}
\includegraphics[width=5cm,scale=0.5,angle=-90,clip]{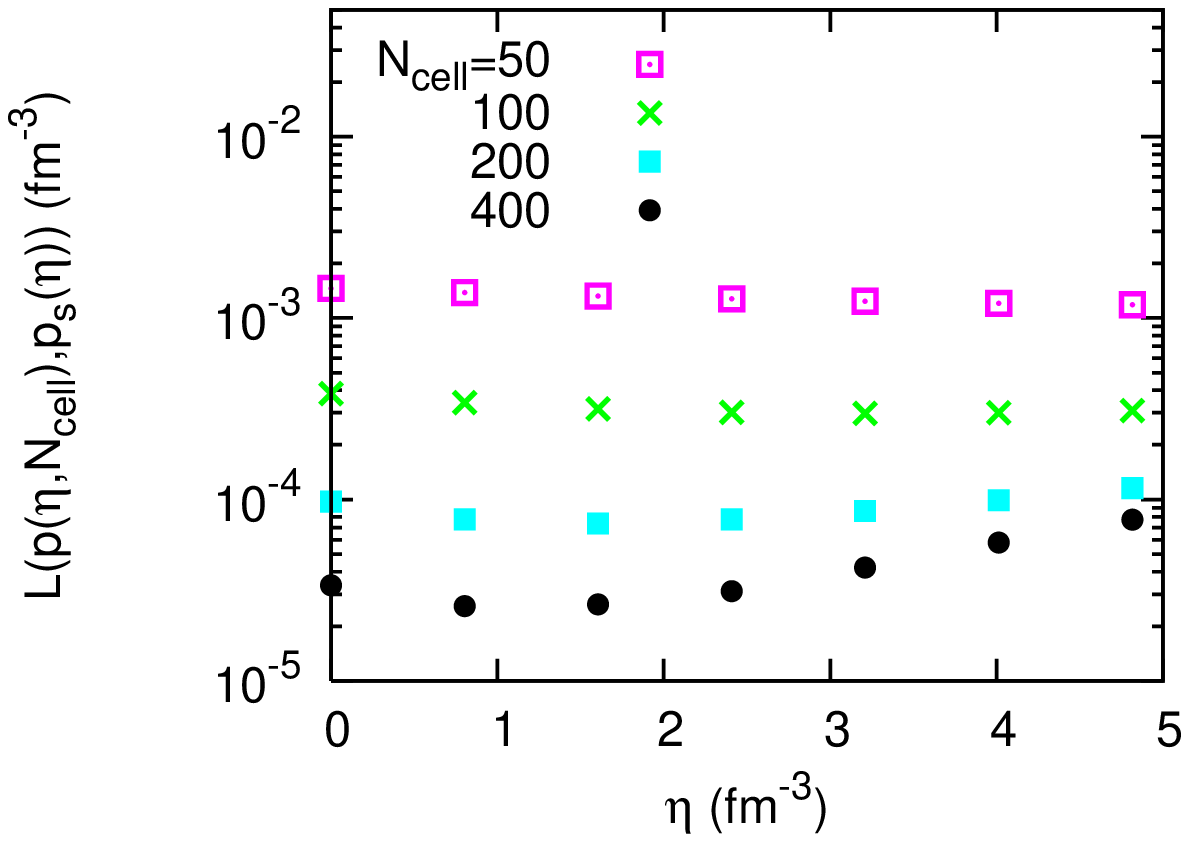}
\caption{
(Color online.)
(Left) L1 norm $L(p(\eta,N_{\rm cell}),p_{\rm s})$ as a function of physical viscosity $\eta$.
The dotted line stands for the L1 norm $L_{\rm lin}(\eta)$ in the linear analysis.
(Right) L1 norm $L(p(\eta,N_{\rm cell}),p_{\rm s}(\eta))$ as a function of physical viscosity $\eta$.
}
\label{fig:attenuation}
\end{figure}

\subsubsection{Sound wave damping by physical viscosity}
In Fig.~\ref{fig:attenuation}, we show the numerical results of sound wave propagation for causal viscous hydrodynamics with the free gas EoS.
In this calculation, we choose $\zeta=\sigma=0$ and $\tau_{\eta}= 10\eta/sT$ and set an initial condition by Eq.~\eqref{eq:init}.
In the left panel, we show our numerical result of the sound wave attenuation due to both physical and numerical viscosities by calculating
\begin{eqnarray}
\label{eq:L1norm2}
L(p(\eta,N_{\rm cell}),p_{\rm s})= 
\sum_{i=1}^{N_{\rm cell}}\mid p(x_i,\lambda/c_{\rm s0};\eta,N_{\rm cell}) - p_{\rm s}(x_i,\lambda/c_{\rm s0})\mid\frac{\lambda}{N_{\rm cell}}.
\end{eqnarray}
It shows that $L(p(\eta,N_{\rm cell}),p_{\rm s})$ converges to $L_{\rm lin}(\eta)$ with larger $N_{\rm cell}$ at fixed $\eta$ and the convergence is faster at larger $\eta$.
This tendency is due to the numerical dissipation $\eta_{\rm num}(\Delta x)$;
when $\eta_{\rm num}(\Delta x)\ll\eta$, the discretization effect is expected to be overwhelmed by the physical viscosity.
In order to disentangle the physical and numerical viscosities, we calculate the following L1 norm:
\begin{eqnarray}
\label{eq:L1norm3}
L(p(\eta,N_{\rm cell}),p_{\rm s}(\eta))= 
\sum_{i=1}^{N_{\rm cell}}\mid p(x_i,\lambda/c_{\rm s0};\eta,N_{\rm cell}) - p_{\rm s}(x_i,\lambda/c_{\rm s0};\eta)\mid\frac{\lambda}{N_{\rm cell}},
\end{eqnarray}
which eliminates the contribution from sound wave attenuation due to the physical viscosity.
The result is shown in the right panel.
We find that $L(p(\eta,N_{\rm cell}),p_{\rm s}(\eta))$ does not depend much on the physical viscosity $\eta$.
This indicates that the numerical dissipation $\eta_{\rm num}\approx \left[c_{\rm s0}(e_0+p_0)/\lambda\right]\cdot (\Delta x)^2$ gives a universal estimate of the numerical dissipation of our scheme in the presence of the physical viscosity.
\footnote{
The increase of $L(p(\eta,N_{\rm cell}),p_{\rm s}(\eta))$ with $N_{\rm cell}=400$ at large $\eta>2 \ {\rm fm}^{-3}$ is due to the limitation of Eq.~\eqref{eq:dispersion}.
The dispersion relation has higher-order contributions
$\omega = \pm c_{\rm s0} k - i\gamma k^2 \pm (\gamma/c_{\rm s0})(c_{\rm s0}^2\tau_{\eta}-\gamma/2)k^3 + {\mathcal O}(k^4)$ \cite{Baier:2007ix}.
At $t=\lambda/c_{\rm s0}$, the third-order term gives $\epsilon\sim {\mathcal O(10^{-4})}\cdot(\eta/{\rm fm^{-3}})^2$ correction to the damped sound wave with Eq.~\eqref{eq:dispersion}, which therefore is different from the exact solution by L1 norm $\delta L\sim \delta p\cdot \epsilon\cdot\lambda\sim {\mathcal O(10^{-5})}\cdot(\eta/{\rm fm^{-3}})^2 \ [{\rm fm}^{-3}]$.
Due to this difference, the L1 norm equation \eqref{eq:L1norm3} saturates at $L>\delta L$ with large $N_{\rm cell}$.
}

In our numerical scheme, the numerical dissipation is $\eta_{\rm num}\approx \left[c_{\rm s}(e+p)/\lambda\right]\cdot (\Delta x)^2=\left(c_{\rm s}sT/\lambda\right)\cdot (\Delta x)^2$, where $s$ denotes the entropy density.
Since we are interested in physical viscosity of $\eta\approx(0.1$-$1)s$, the condition $(\eta_{\rm num}/\eta) \approx [c_{\rm s}T/(0.1$-$1)\lambda]\cdot (\Delta x)^2\ll 1$ gives $\Delta x \ll 0.8$-$2.6 \ {\rm fm}$ at $T=500 \ {\rm MeV}$ and with $\lambda=10 \ {\rm fm}$, which are the typical temperature and system length scale at the relativistic heavy-ion collisions.
This condition becomes more severe at higher temperature or when finer structure is of interest.
We emphasize that appropriate fine grid size calculation is indispensable for any physical observables in heavy-ion collisions, 
to discuss the value of physical viscosity from  comparison with experimental data. 

\subsection{Shock tube problem}
The shock tube problem is analytically solvable for a perfect fluid with the free gas EoS. 
It provides an important test for measuring the performance and accuracy of different numerical schemes. 
To compare our numerical algorithm to other numerical schemes (SHASTA, NT, KT schemes) and the analytical solution \cite{Niemi}, we start the test calculation with the same initial conditions as those of Ref. \cite{Molnar:2009tx}. 
The initial temperature on the left is $T_L=400 $ MeV, and that on the right is $T_R=200$ MeV. 
In the calculation we employ the free gas EoS. 
The spatial cell size and the Courant number are set to be $\Delta x =0.1$ and 
$\lambda=0.4$
\footnote{
In our algorithm the Courant number is 
determined based on the Courant-Friedrichs-Lewy (CFL) Condition, 
which produced high-precision calculations (Appendix~\ref{sec:numerical_algorithm}).
},
respectively. 
Because a numerical calculation with fine-enough grid and time step should converge on the analytical solution,  the same discretization for spatial grid size and time step is important for accuracy testing of numerical methods.  

\begin{figure}[t!]
\includegraphics[width=7cm,scale=0.5,clip]{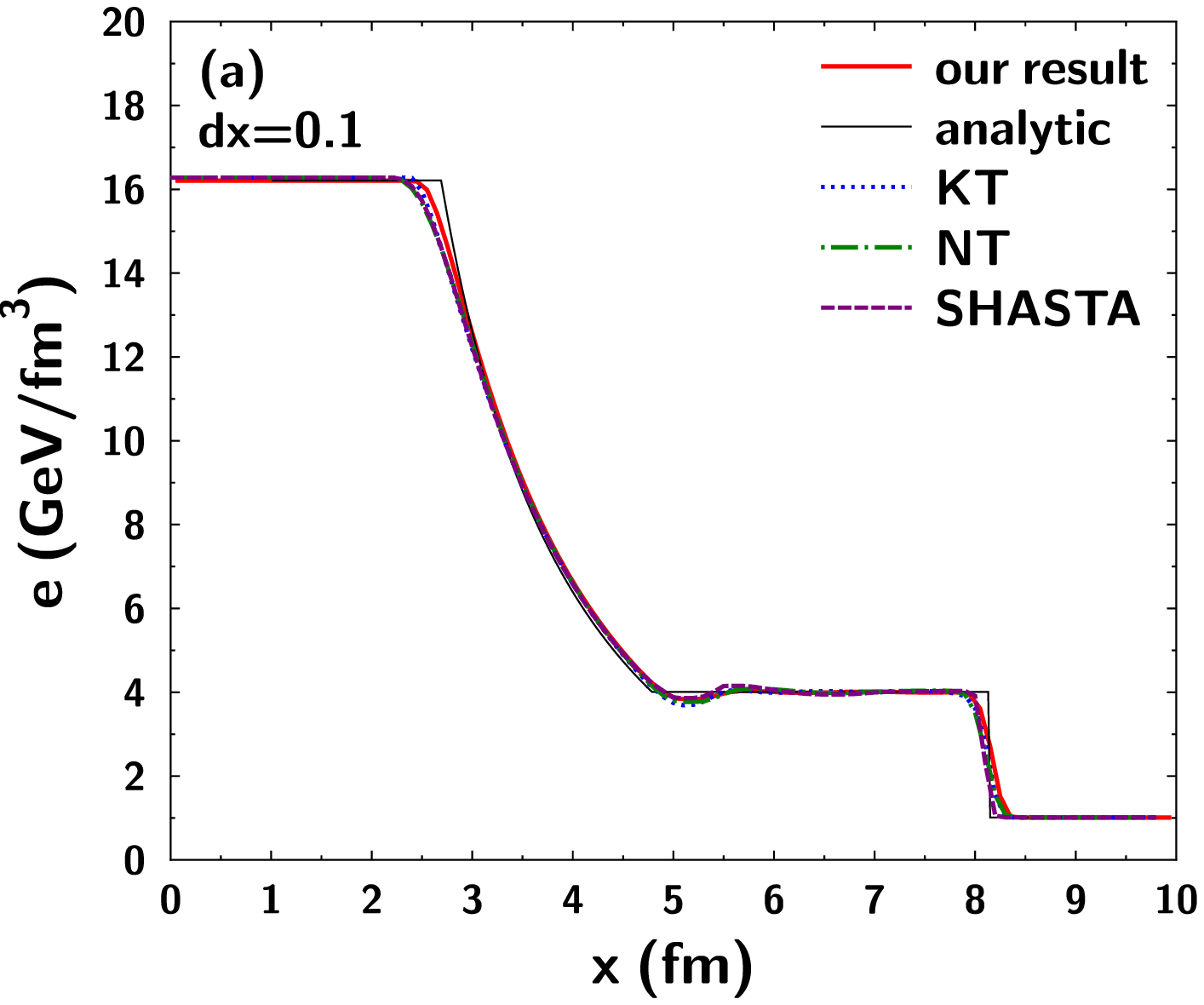}
\includegraphics[width=7cm,scale=0.5,clip]{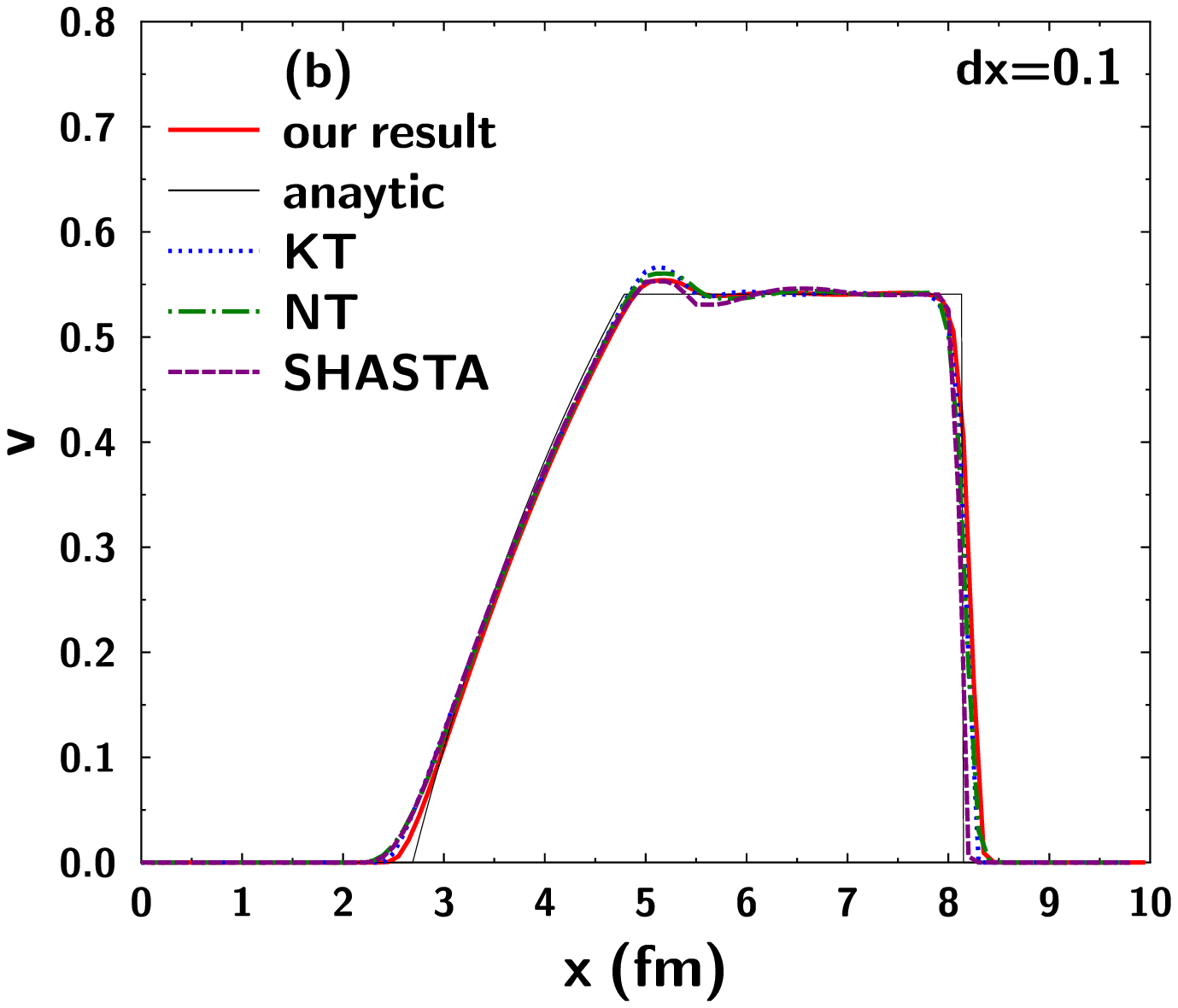}
\includegraphics[width=7cm,scale=0.5,clip]{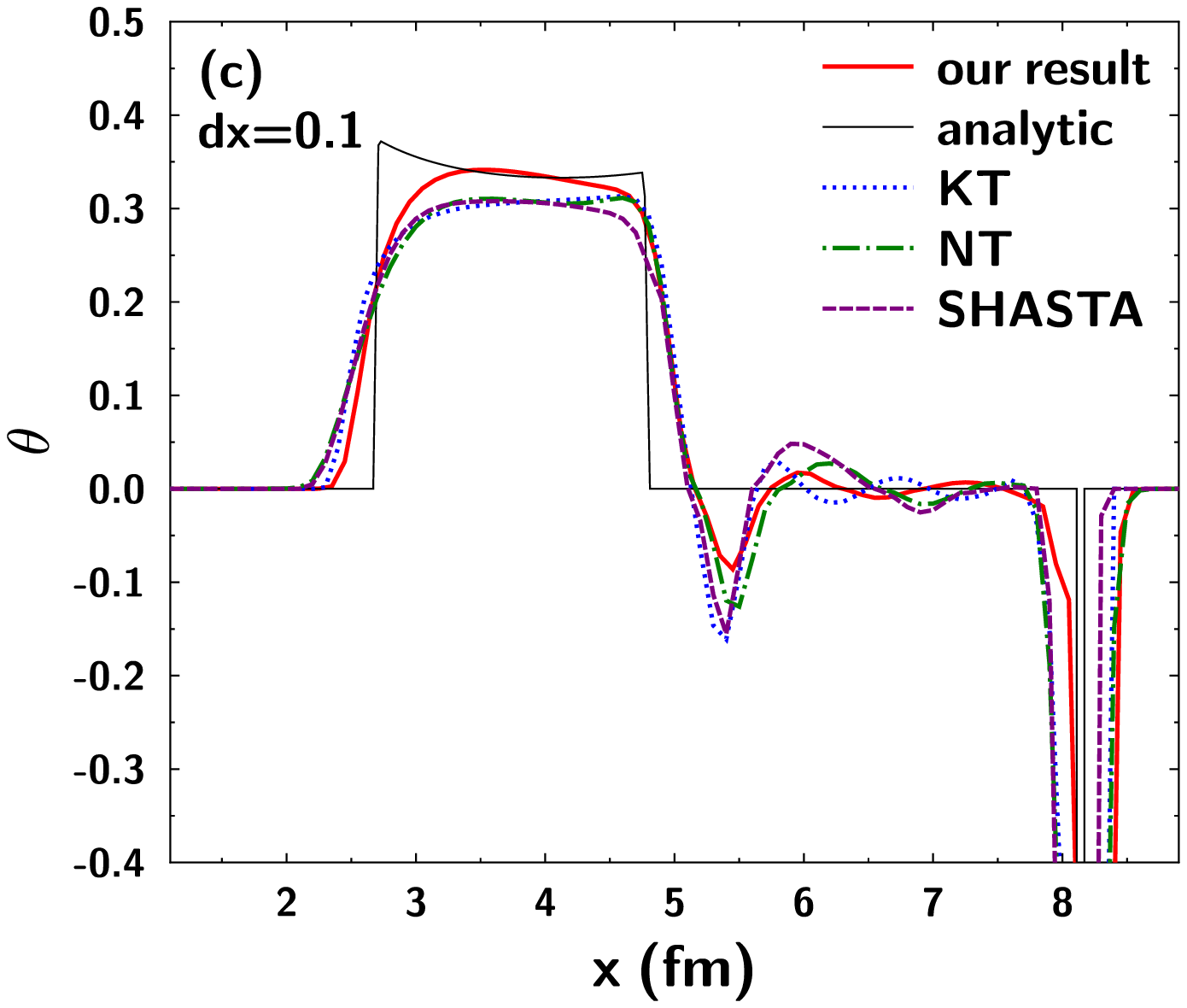}
\caption{
(Color online.)
The analytic (thine line) and numerical solutions of the relativistic Riemann 
problem on a grid with $N_x=100$ cells with $\Delta x=0.1$ fm, after $N_t=100$ time 
steps at $t= 4$ fm/c. (a) The energy density distribution $e$, (b) the velocity $v$, 
(c) the invariant expansion rate $\theta$ with our algorithm (solid line) KT (dotted line), NT (dashed-dotted line), 
and SHASTA (dashed line). 
}
\label{fig:shock-tube-test}
\end{figure}

Fig. \ref{fig:shock-tube-test} shows the energy density distribution, the velocity and 
the invariant expansion rate $\theta=\partial_\mu u^\mu$ with our algorithm, KT, NT, and SHASTA, together with 
the analytical solution for an ideal fluid.  
For these values, KT, NT, and SHASTA algorithms reproduce the analytical solution with almost the same 
accuracy and numerical artifacts. The difference between the analytical solution and 
numerical calculations indicates existence of numerical dissipation in numerical schemes. 
It is worth noting that, our numerical results are closer to the analytical solution, especially at $x=3$ fm compared to KT, NT, and SHASTA algorithms, which suggests that our algorithm contains less numerical dissipation. 
This tendency appears clearly in the invariant expansion rate $\theta$ in Fig.\ \ref{fig:shock-tube-test}. 
Moreover, only our numerical scheme follows the shape of the analytical solution from $x=3$ fm to $x=5$ fm. 
Numerical dissipation is indispensable for the stability of numerical calculations of the relativistic hydrodynamical equation. However, too much numerical dissipation smears numerical results and leads a solution far off from the analytical one. 

\begin{figure}[t!]
\includegraphics[width=5cm,scale=0.5,angle=-90,clip]{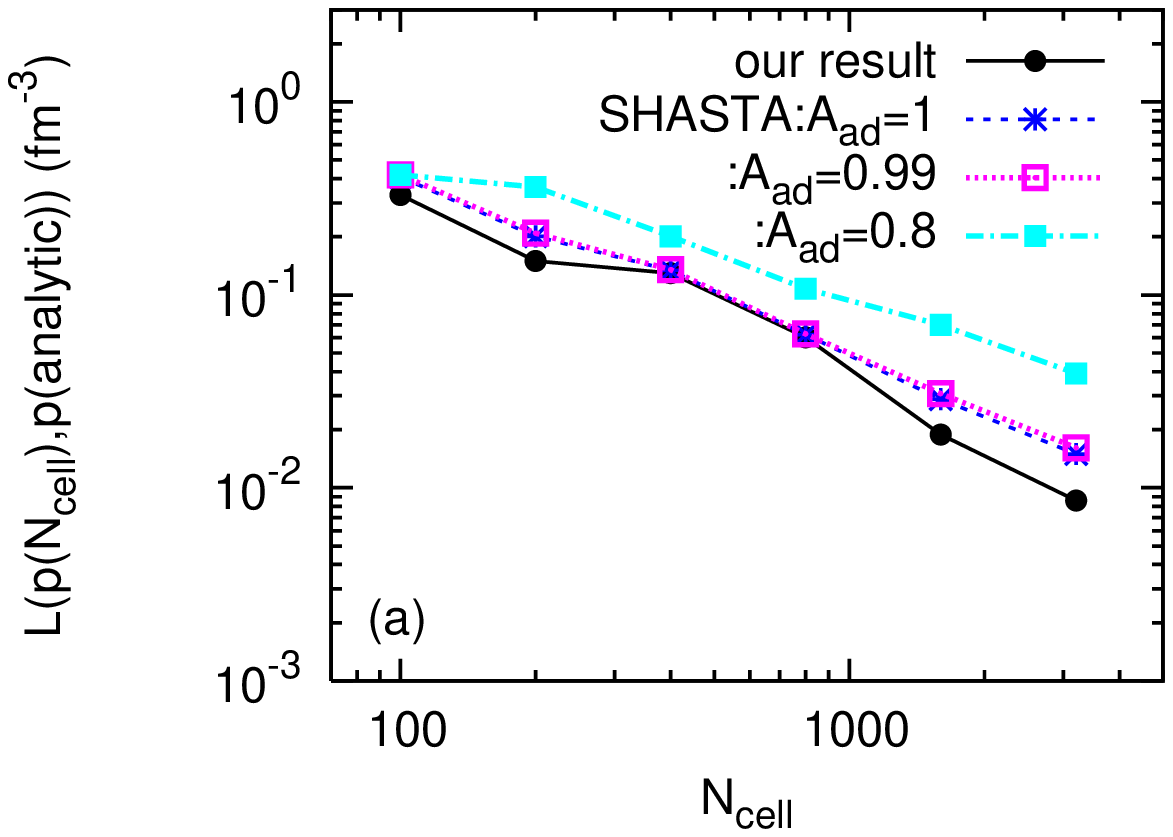}
\includegraphics[width=5cm,scale=0.5,angle=-90,clip]{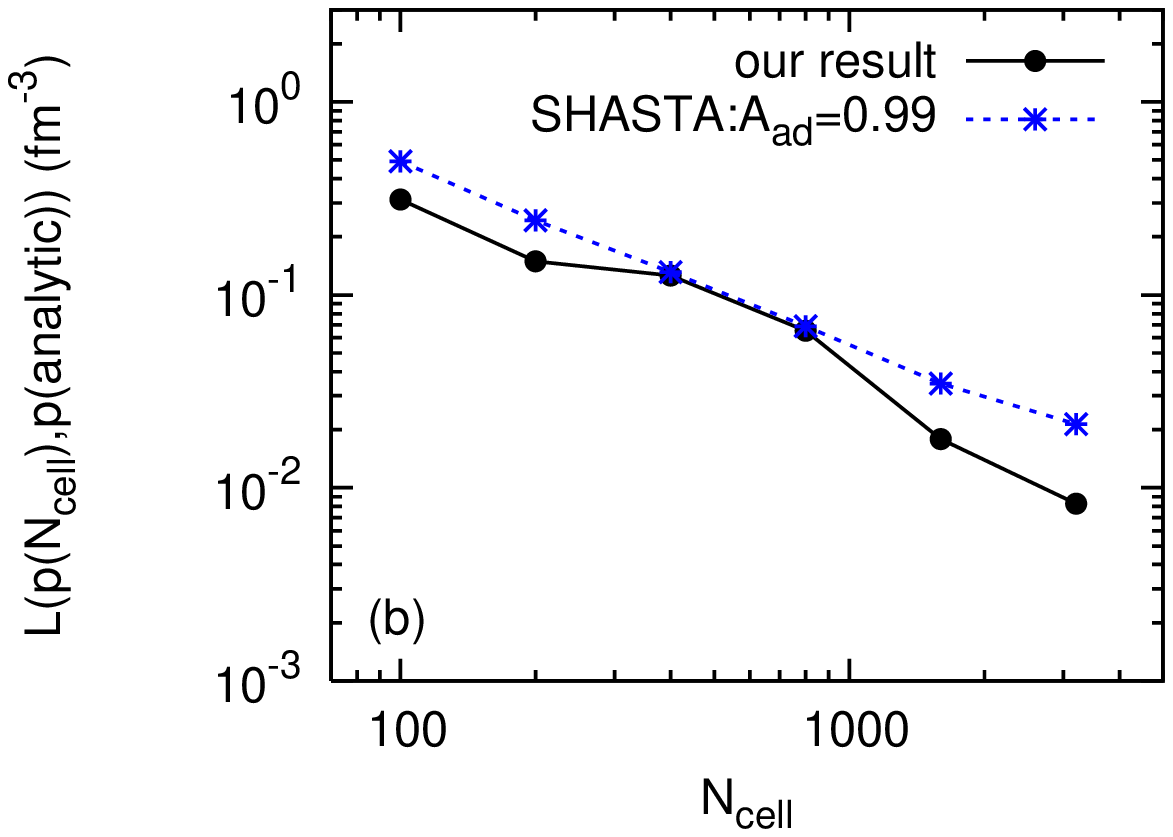}
\includegraphics[width=5cm,scale=0.5,angle=-90,clip]{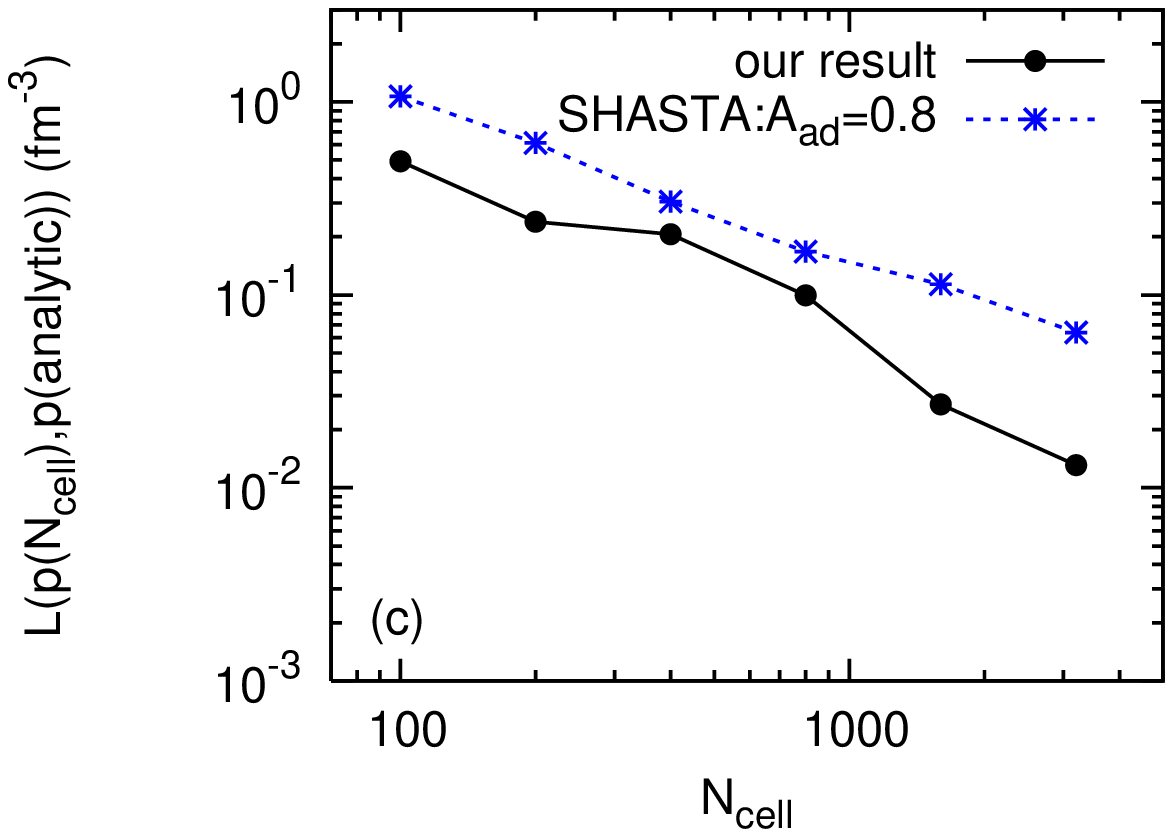}
\caption{
(Color online.)
L1 norm errors for the shock tube problems in our algorithm and the SHASTA scheme. 
(a) $(T_L, T_R)$ =(400 MeV, 200 MeV), (b) $(T_L,  T_R)$ =(400 MeV, 172 MeV)  and 
(c) $(T_L, T_R) $=(450 MeV, 170 MeV). 
}
\label{fig:l1norm-shock-tube}
\end{figure}

We evaluate the L1 norm for the shock tube problems using our algorithm and the SHASTA scheme which is often used in 
hydrodynamic models applied to high-energy heavy-ion collisions. The CFL number is set to be 0.4 in the following L1 norm calculation. 
In Fig.\ref{fig:l1norm-shock-tube} (a) the L1 norm errors  
of the SHASTA scheme with $A_{\rm ad}=1$, 0.99 and 0.8 and our algorithm are shown. Here the initial temperatures on the left and the right 
are as the same as ones in Fig.~\ref{fig:shock-tube-test}.  We find that the L1 norm of our algorithm is smaller 
than that of the SHASTA scheme for each $N_{\rm cell}$, which suggests that our algorithm has smaller numerical dissipation 
compared to the SHASTA. 
The difference of the L1 norm between our algorithm and the SHASTA  scheme becomes large, as the value of $A_{\rm ad}$ decreases. 

We find that the SHASTA scheme with $A_{\rm ad}=1$ becomes unstable, if the temperature difference between the left and 
the right becomes large. 
For example, in the case of the initial temperature on the left $T_L=400$ MeV and that on the right $T_R=172$ MeV, 
the calculation with the SHASTA with $A_{\rm ad}=1$ does not work. To stabilize the numerical calculation with the SHASTA, 
we change $A_{\rm ad}$ from 1 to 0.99, which means introduction of additional numerical dissipation to the SHASTA.  
On the other hand, our algorithm is stable with the initial temperatures  without any additional numerical dissipation. 
This difference appears in the value of the L1 norm.  In Fig.~\ref{fig:l1norm-shock-tube} (b)  
we can see that the difference between the L1 norm of our algorithm and that of the SHASTA scheme becomes larger, compared to 
the difference between them in  Fig.~\ref{fig:l1norm-shock-tube} (a).
Furthermore  in the case of $(T_L, T_R) =$ (450  MeV, 170 MeV),  $A_{\rm ad}$ is set to be 0.8 for stability of the numerical calculation 
in the SHASTA. 
Fig.~\ref{fig:l1norm-shock-tube} (c) indicates that the SHASTA algorithm has large numerical dissipation compared to our algorithm. 

In analyses of high-energy heavy-ion collisions with hydrodynamic model, such a temperature difference between cells can be realized.  
For instance, the maximum value of initial temperature for Au+Au $\sqrt{s_{NN}}=200$ GeV collisions at RHIC is estimated to be 300 - 600 MeV \cite{Nonaka:2012qw}.  
In the heavy-ion collisions at LHC higher temperature is achieved.  
On the other hand, we can utilize the hydrodynamic picture if the temperature of the system is above $T \sim150$ MeV \cite{Nonaka:2012qw}.  
Therefore, the temperature fluctuations between $T=450$ MeV and $T=170$ MeV which is shown in the previous shock tube problems 
can exist in an initial temperature distribution for the high-energy heavy-ion collisions. This fact suggests that the numerical 
scheme that is stable for strong shock wave with small numerical dissipation is more suitable for investigation of physics of 
high-energy heavy-ion collisions. Our  algorithm has an advantage over the  SHASTA scheme on this point.

\begin{figure}[t!]
\includegraphics[width=7cm,scale=0.5,clip]{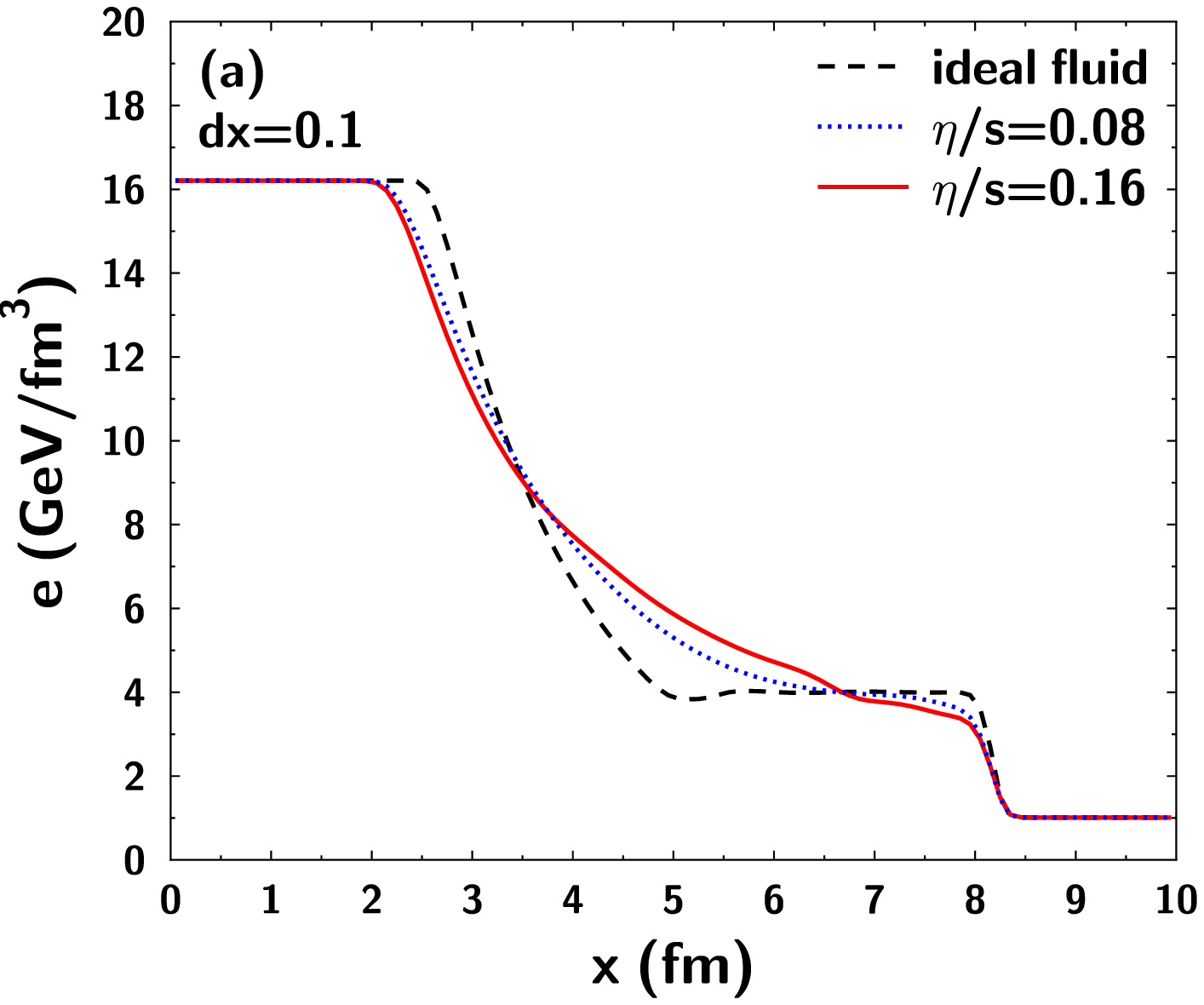}
\includegraphics[width=7cm,scale=0.5,clip]{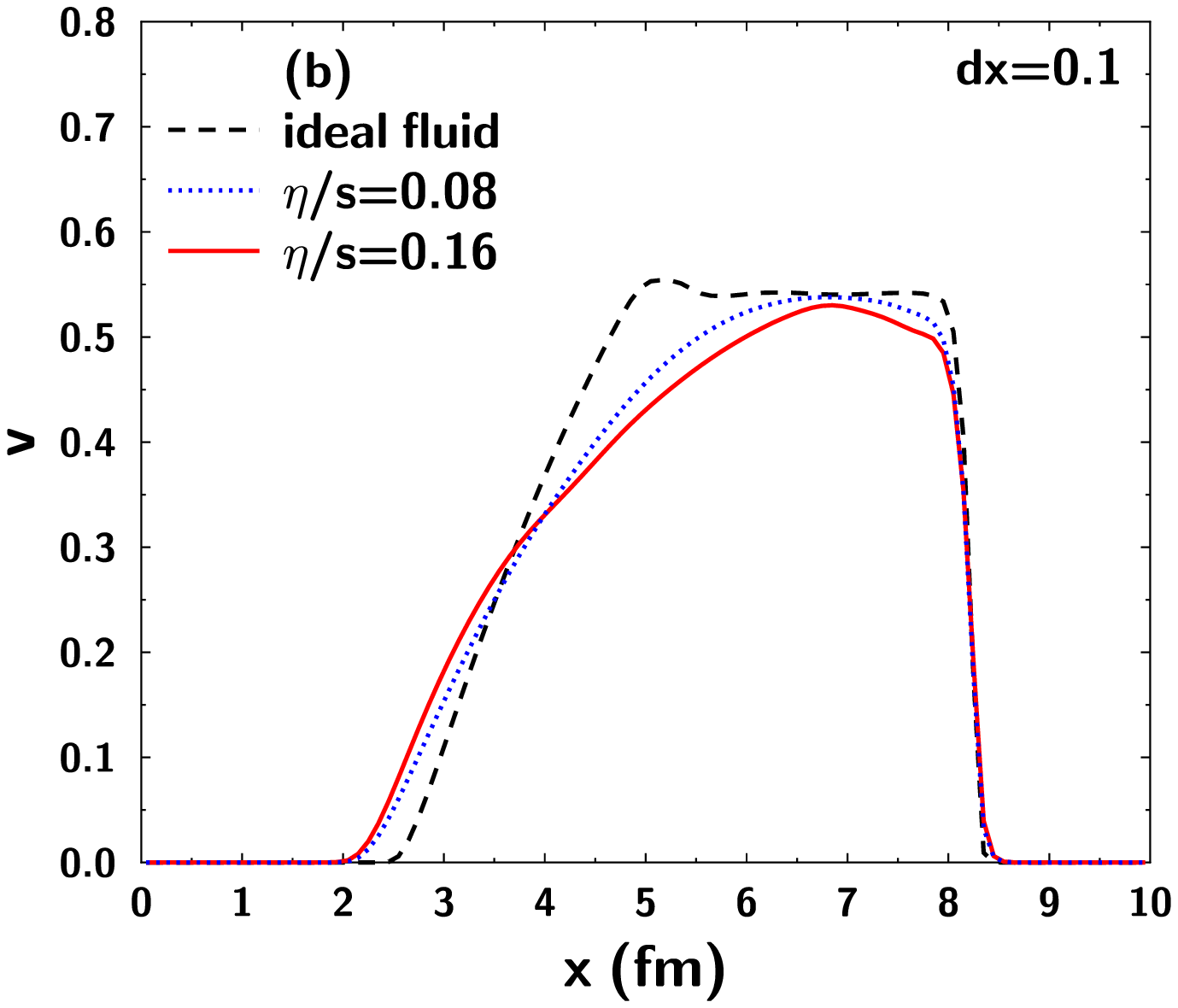}
\includegraphics[width=7cm,scale=0.5,clip]{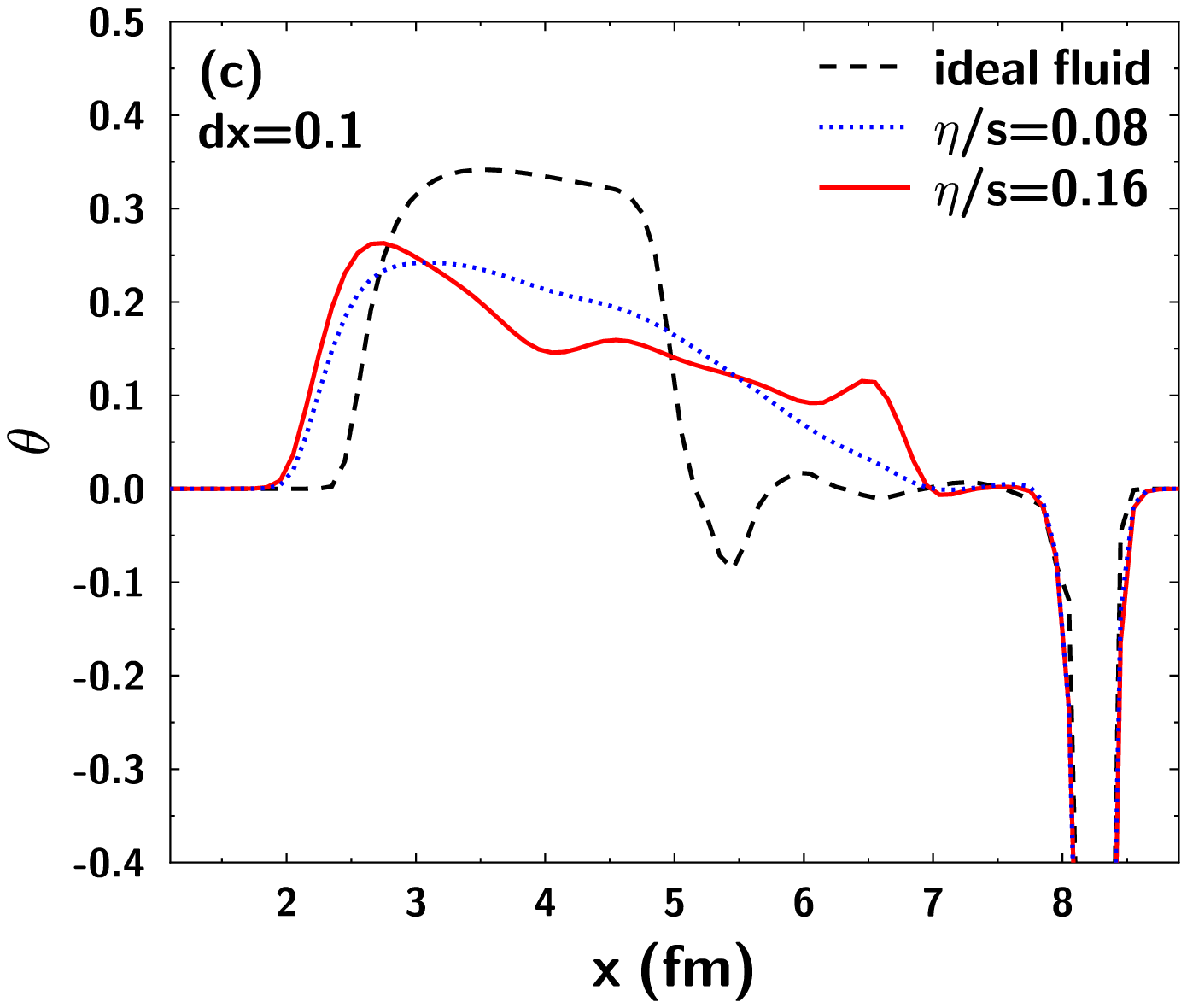}
\caption{
(Color online.)
Shear viscosity dependence of (a) energy density $e$, 
(b) velocity $v$ and (c) invariant expansion rate $\theta$.
}
\label{fig:shock-tube-test-eta}
\end{figure}

\begin{figure}[t]
\includegraphics[width=7cm,scale=0.5,clip]{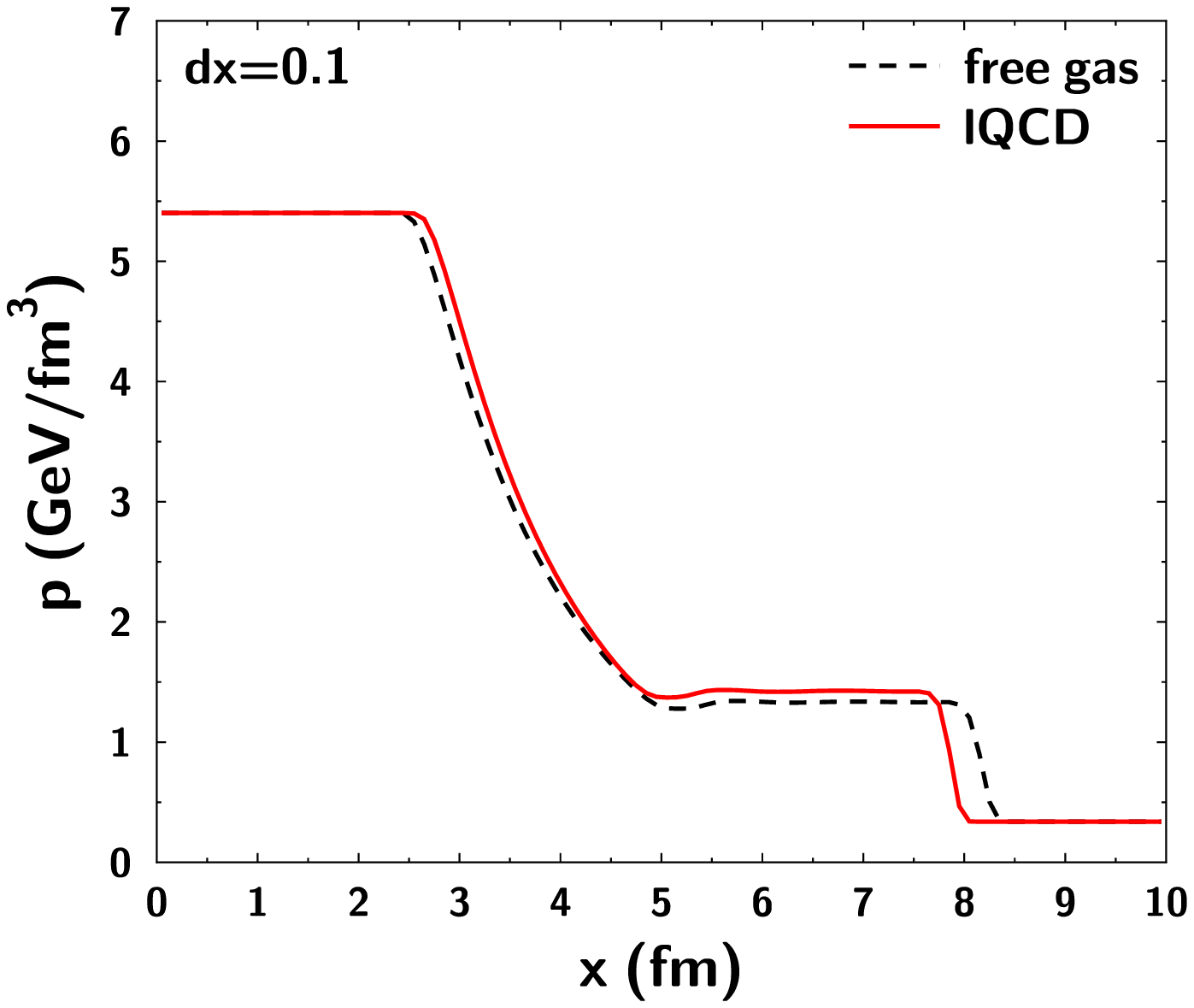}
\includegraphics[width=7cm,scale=0.5,clip]{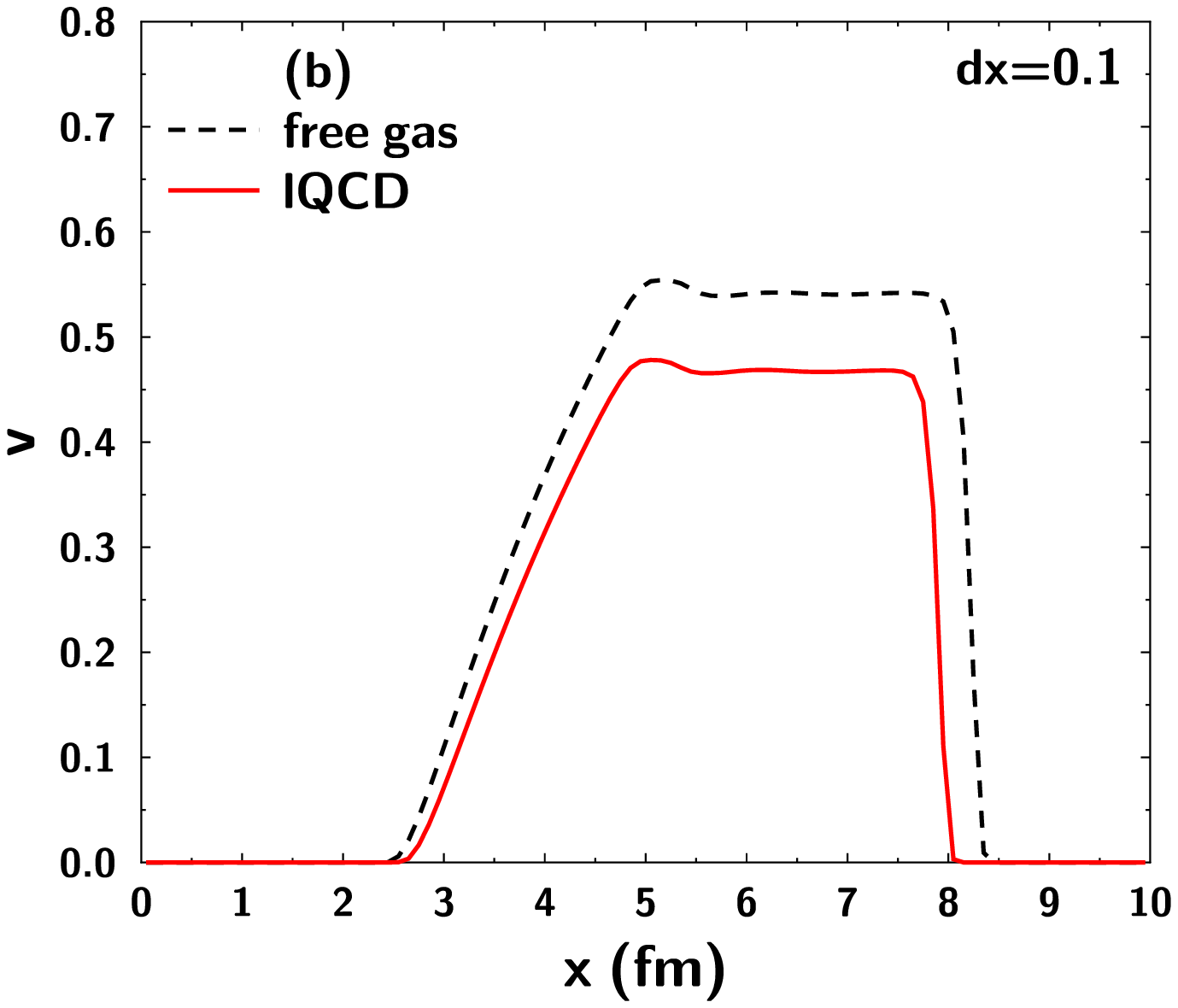}
\includegraphics[width=7cm,scale=0.5,clip]{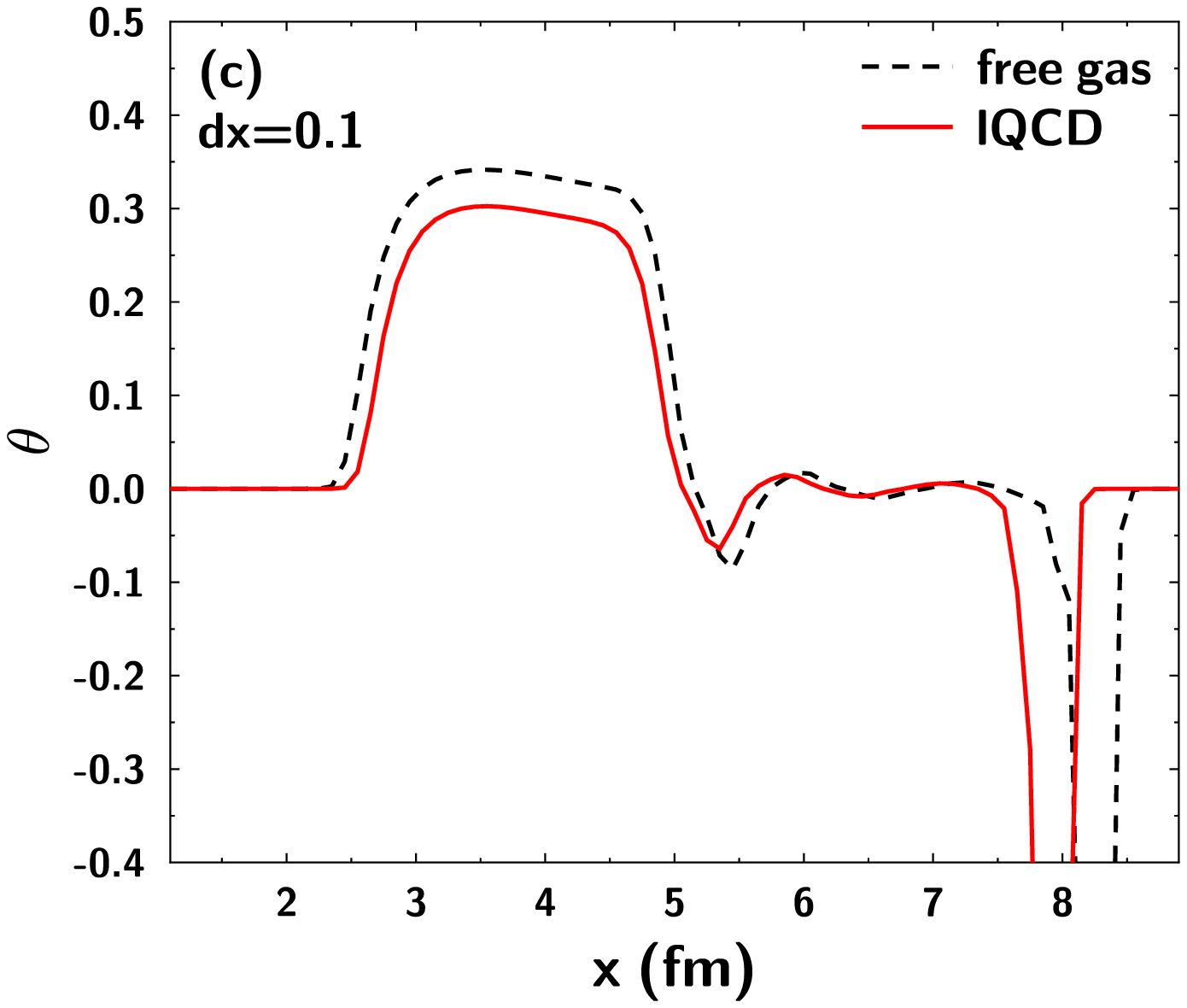}
\caption{
(Color online.) 
EoS dependence of (a) pressure $p$,  
(b) velocity $v$ and (c) invariant expansion rate $\theta$.
}
\label{fig:shock-tube-test-qcd}
\end{figure}

If fine-enough cell size is utilized in numerical calculation, 
the distinction among different algorithms becomes small,  because numerical solutions should converge to the analytical one. 
However, the speed of convergence to the analytical solution varies among different numerical schemes. 
For example, to analyze the higher harmonics induced by event-by-event fluctuations in experiments,  
we need to carry out numerical calculations with fluctuating initial conditions, which 
indicates that we reconcile a numerical calculation on coarser grids under 
current computational resources.
According to the physics application of hydrodynamics, we need to choose an appropriate 
numerical method for solving the relativistic hydrodynamics equation. 
Besides, in relativistic heavy-ion collisions, one of the interesting and important topics is 
investigating bulk properties of the QGP, such as its transport coefficients. 
To evaluate the physical viscosities of QGP from analyses of experimental data based on 
hydrodynamic models, we need to control the numerical dissipation. 
The difficulty of distinguishing between the physical viscosity and the numerical dissipation  was discussed in Ref.\ \cite{Molnar:2009tx}. For investigation of 
physical viscosity of QGP,  the algorithm in which the numerical dissipation is well controlled is indispensable. 

Fig.~\ref{fig:shock-tube-test-eta} shows the shear viscosity dependence of the energy density distribution, velocity and invariant expansion rate. 
At finite shear viscosity, deviation from the result of the ideal fluid becomes large and the shape of distribution is smeared. 
We observe the same tendency in finite bulk viscosity and baryon number conductivity calculation. 

Fig.~\ref{fig:shock-tube-test-qcd} shows the EoS dependence of the pressure
distribution, velocity and invariant expansion rate.
For comparison, the same initial pressure distribution is employed for both cases. 
The fact that the sound velocity of lattice QCD EoS is smaller than that of the free gas EoS (Fig.~\ref{fig:eos}) affects expansion rate. In Fig.~\ref{fig:shock-tube-test-qcd} (c) 
the expansion rate of lattice QCD EoS is smaller that that of the free gas EoS in almost 
everywhere. 
As a consequence, the velocity of lattice QCD EoS is smaller than that of the free gas EoS  (Fig.~\ref{fig:shock-tube-test-qcd} (b))  and 
expansion of the shock wave in pressure distribution is smaller than that of the free gas EoS (Fig.~\ref{fig:shock-tube-test-qcd} (a)) .

\subsection{Blast wave problem}
We solve a (2+1)-dimensional blast wave problem.
The initial pressure and density are uniform, and the initial flow vector is normalized to $v_r$ and points to the center of the system:
\begin{eqnarray}
{\bm V}(x,y,t=0)
= \left(
0,
-\frac{v_r x}{\sqrt{x^2+y^2}},
-\frac{v_r y}{\sqrt{x^2+y^2}},
0,
p_0
\right),
\end{eqnarray}
with $p_0=1 \ {\rm fm^{-4}}$ and $v_r=0.9$.
The system area is a square with $6 \ {\rm fm}\times 6 \ {\rm fm}$ square, we discretize it with 384 points in each direction.

We perform the blast wave simulation in the
(i) ideal and viscous hydrodynamics with free gas EoS and
(ii) ideal and viscous hydrodynamics with lattice QCD EoS.
In viscous hydrodynamic simulations, we choose viscous coefficients $\eta/s=0.1, \ \zeta=0 \ {\rm fm}^{-3}$, baryon number conductivity $\sigma=0 \ {\rm fm}^{-1}$, and relaxation time for the shear mode $\tau_{\eta}=10\eta/sT=1/T$.

\begin{figure}[t!]
\raisebox{2mm}{\includegraphics[width=7cm,scale=0.9,angle=-90,clip]{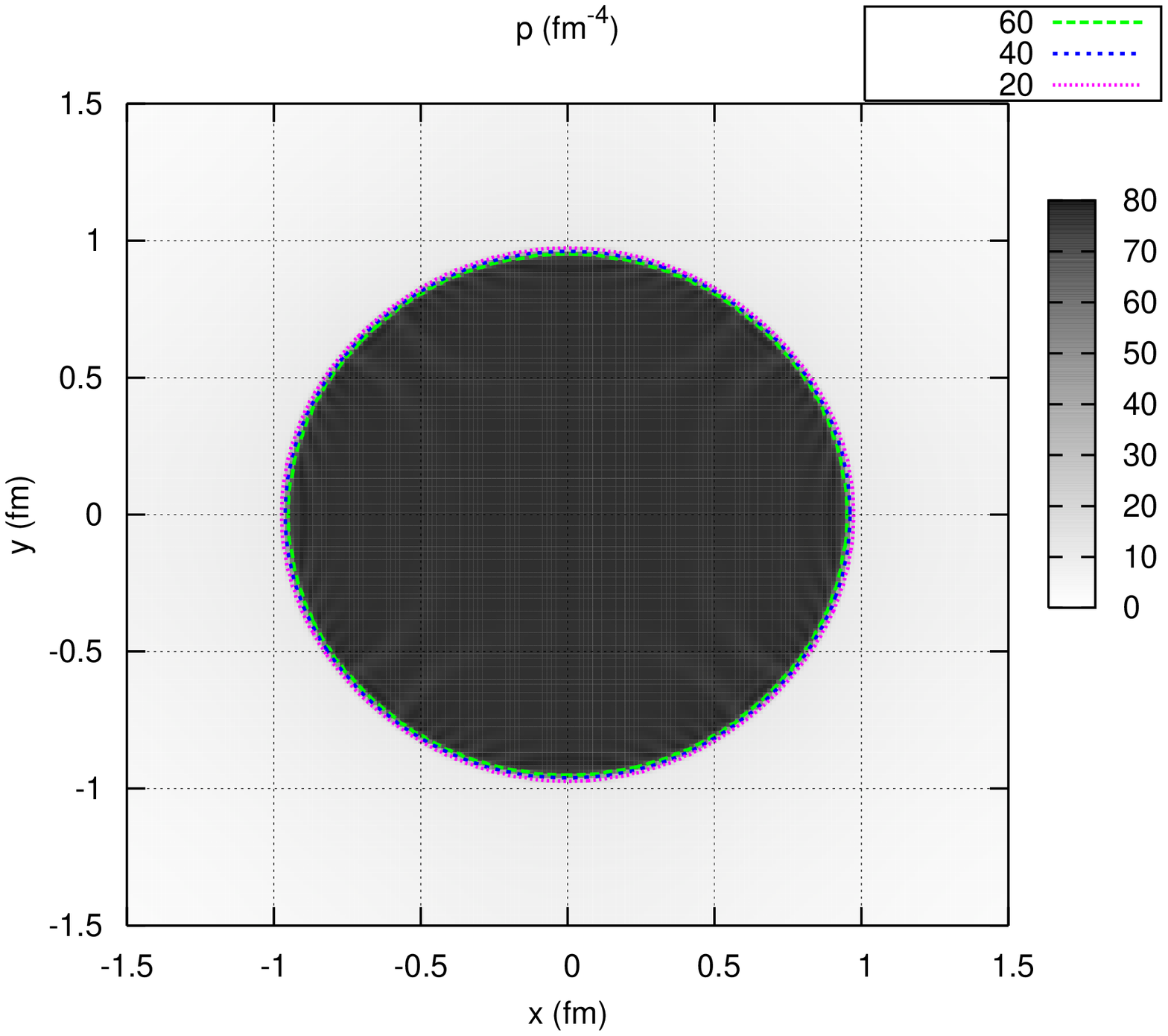}}
\includegraphics[width=7cm,scale=0.9,angle=-90,clip]{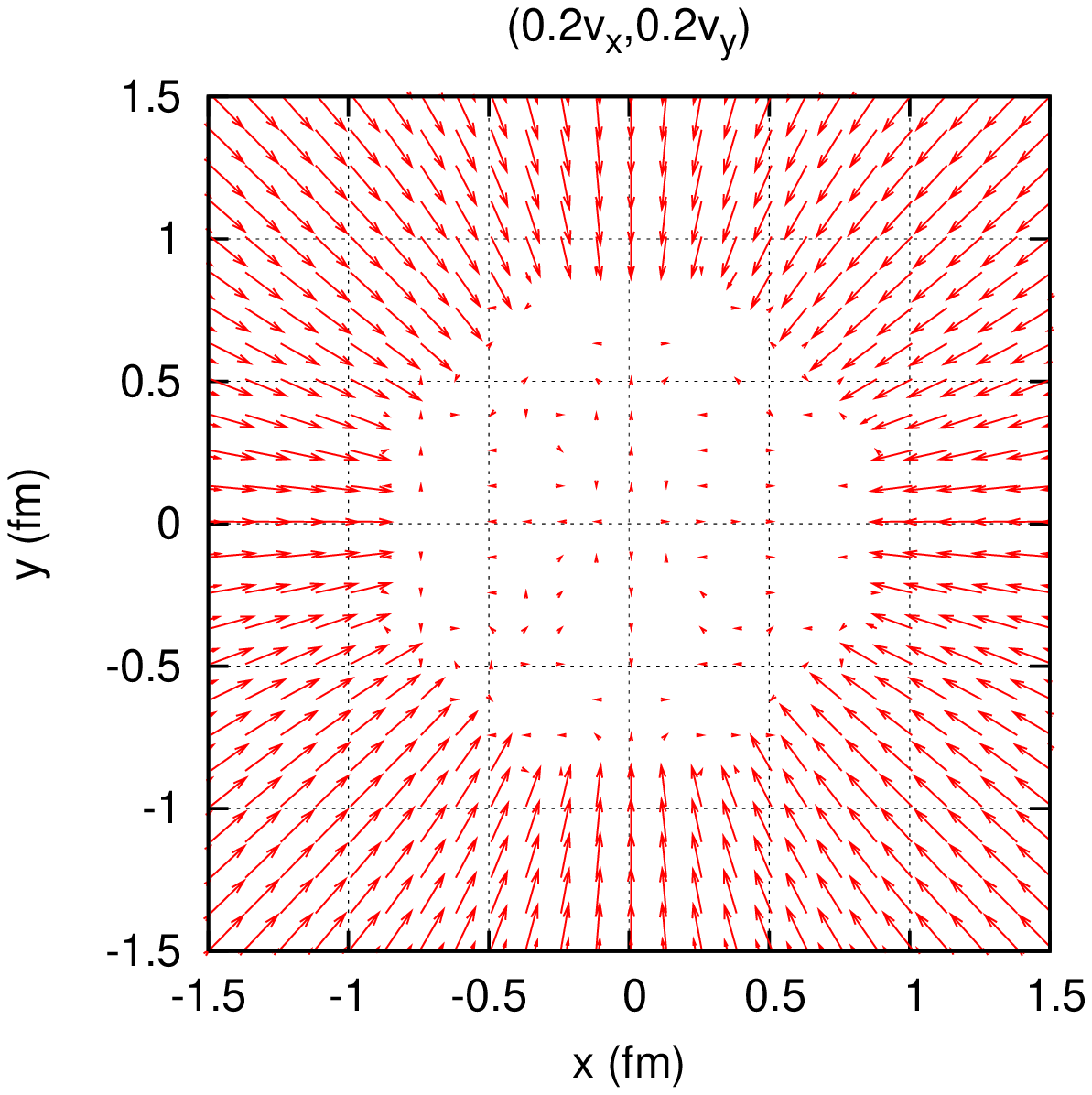}
\includegraphics[width=5.5cm,scale=0.9,angle=-90,clip]{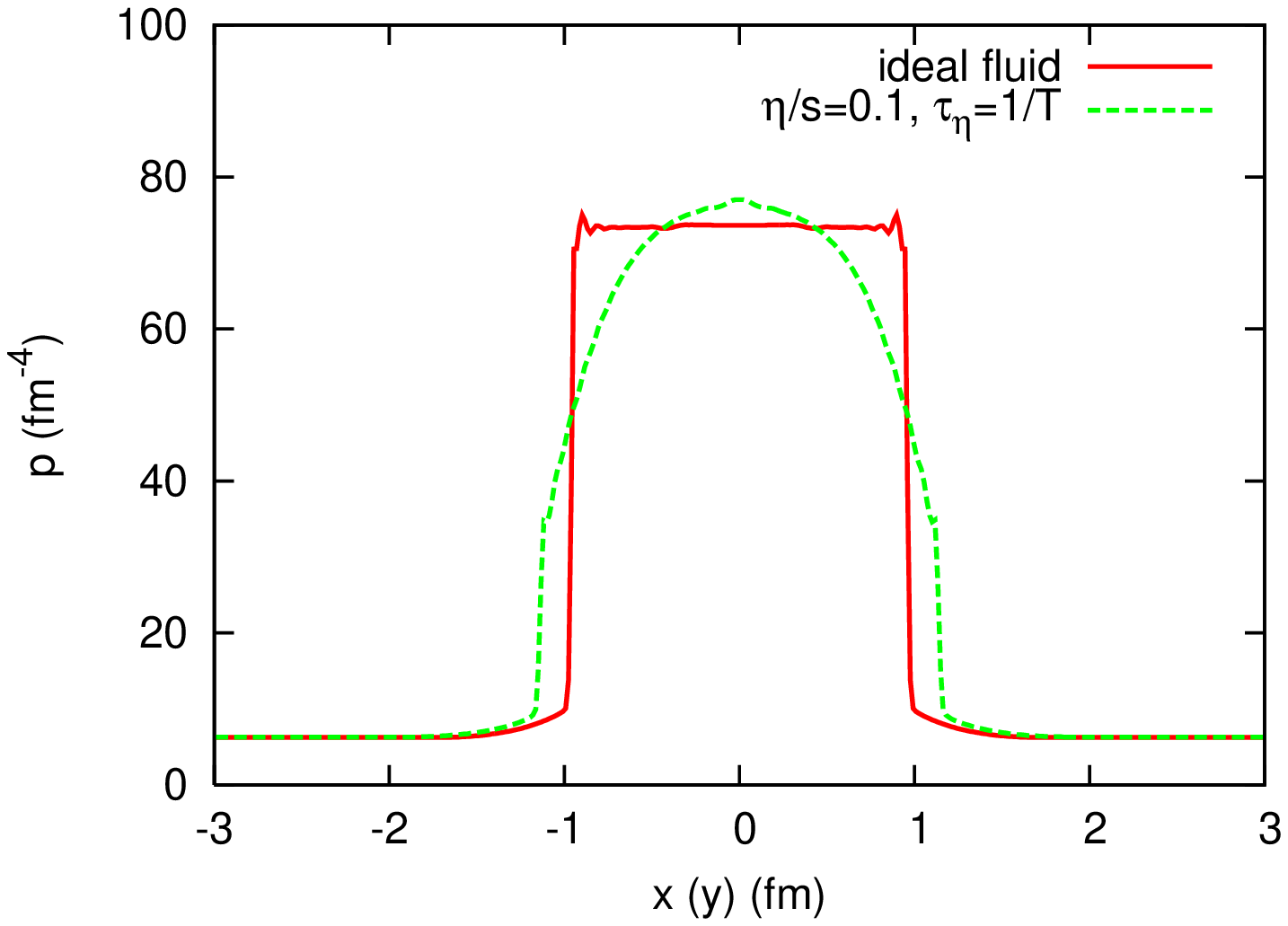}
\includegraphics[width=5.5cm,scale=0.9,angle=-90,clip]{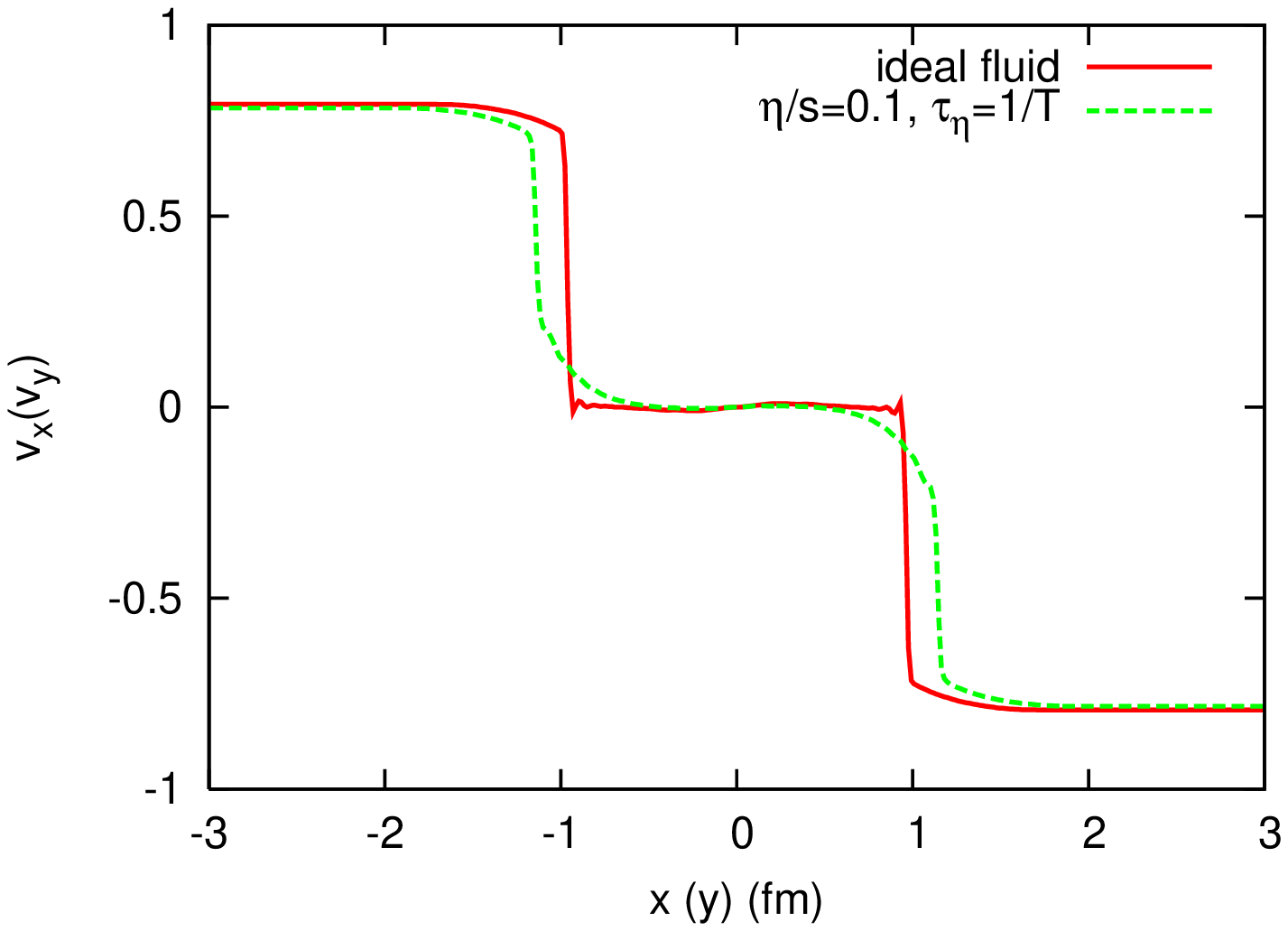}
\caption{
(Color online.)
Simulation of ideal and viscous hydrodynamics with free gas EoS at $t=2.44$ fm (1500 steps).
The upper panels are two-dimensional profiles of (left) pressure and (right) flow velocity for the ideal hydrodynamic simulation.
The lower panels are one-dimensional profiles of (left) pressure and (right) $x(y)$-component of flow velocity at $y(x)=0$ fm for both ideal and viscous hydrodynamic simulations.
The finite viscous coefficient is $\eta/s=0.1$ and the relaxation time for the shear mode is $\tau_{\eta}=1/T$.
}
\label{fig:blast_free}
\end{figure}

In Fig.~\ref{fig:blast_free}, we show the results of simulation (i) at $t=2.44$ fm (1500 steps).
In the upper panels, we plot the pressure and velocity profiles for the ideal hydrodynamic simulation.
Note that the flow velocity field is dimensionless.
At the center, we find a region with high pressure and vanishing flow velocity, which grows in time.
In the lower panels, we show one-dimensional profiles of pressure and $x(y)$-component of flow velocity at $y(x)=0$ fm for both ideal and viscous hydrodynamic simulations.
It is clear that there is a symmetry between $x$ and $y$ directions, which must be realized because of the initial conditions.
We find that the discontinuous change of pressure and flow velocity at 
$\sqrt{x^2+y^2}\approx 1$ fm in the ideal hydrodynamic simulation becomes continuous due to the finite shear viscosity.

\begin{figure}[t!]
\raisebox{2mm}{\includegraphics[width=7cm,scale=0.9,angle=-90,clip]{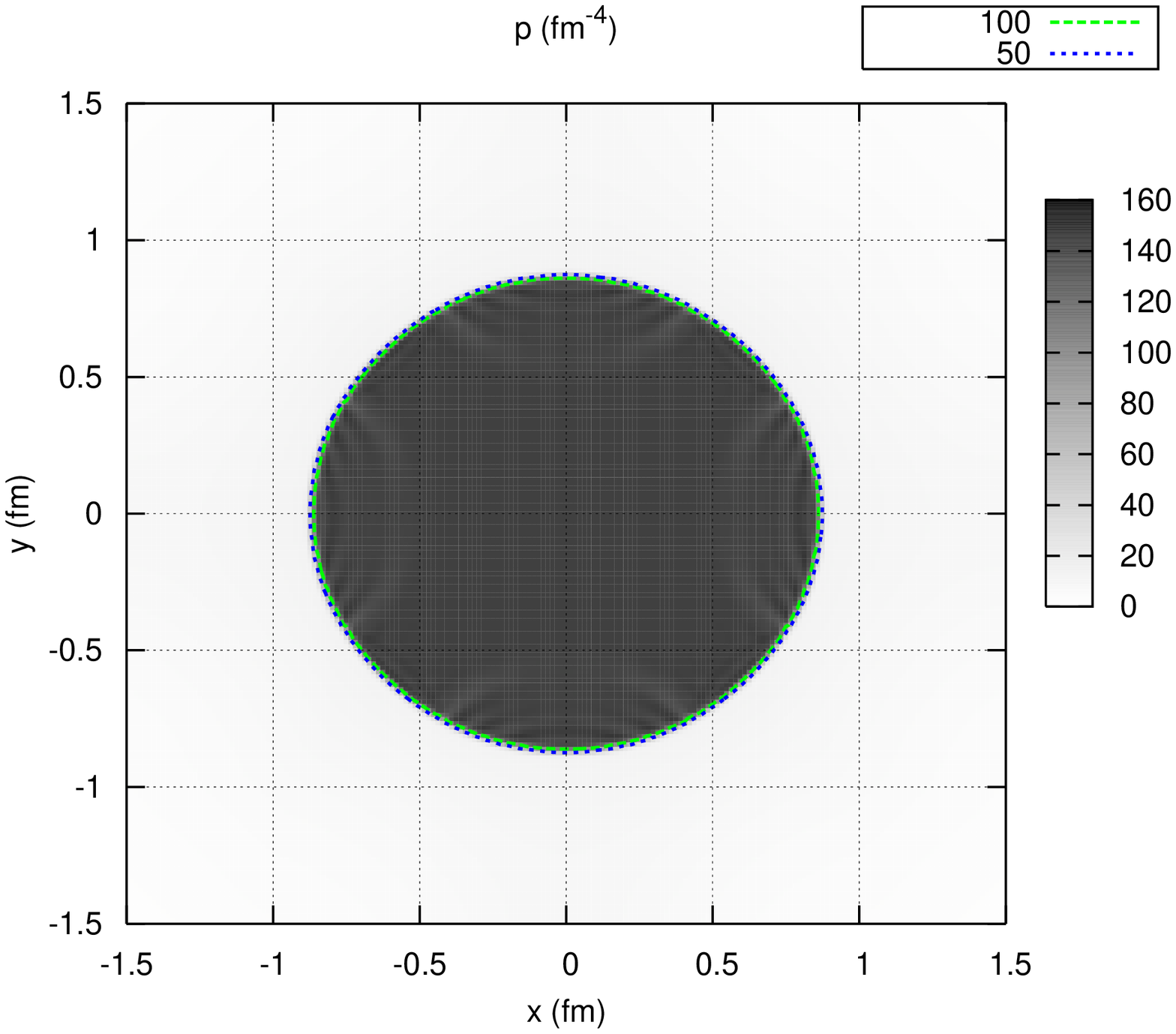}}
\includegraphics[width=7cm,scale=0.9,angle=-90,clip]{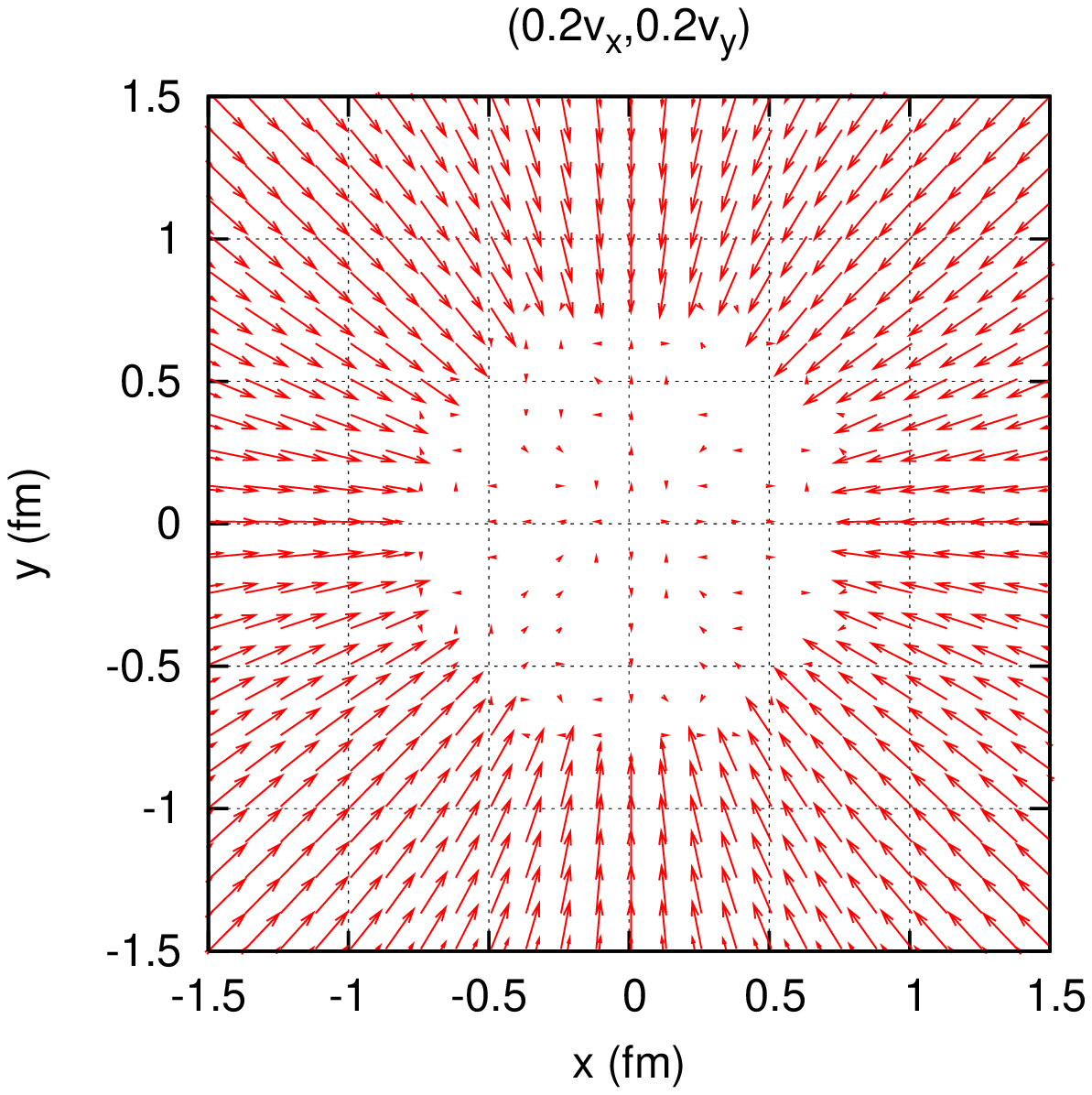}
\includegraphics[width=5.5cm,scale=0.9,angle=-90,clip]{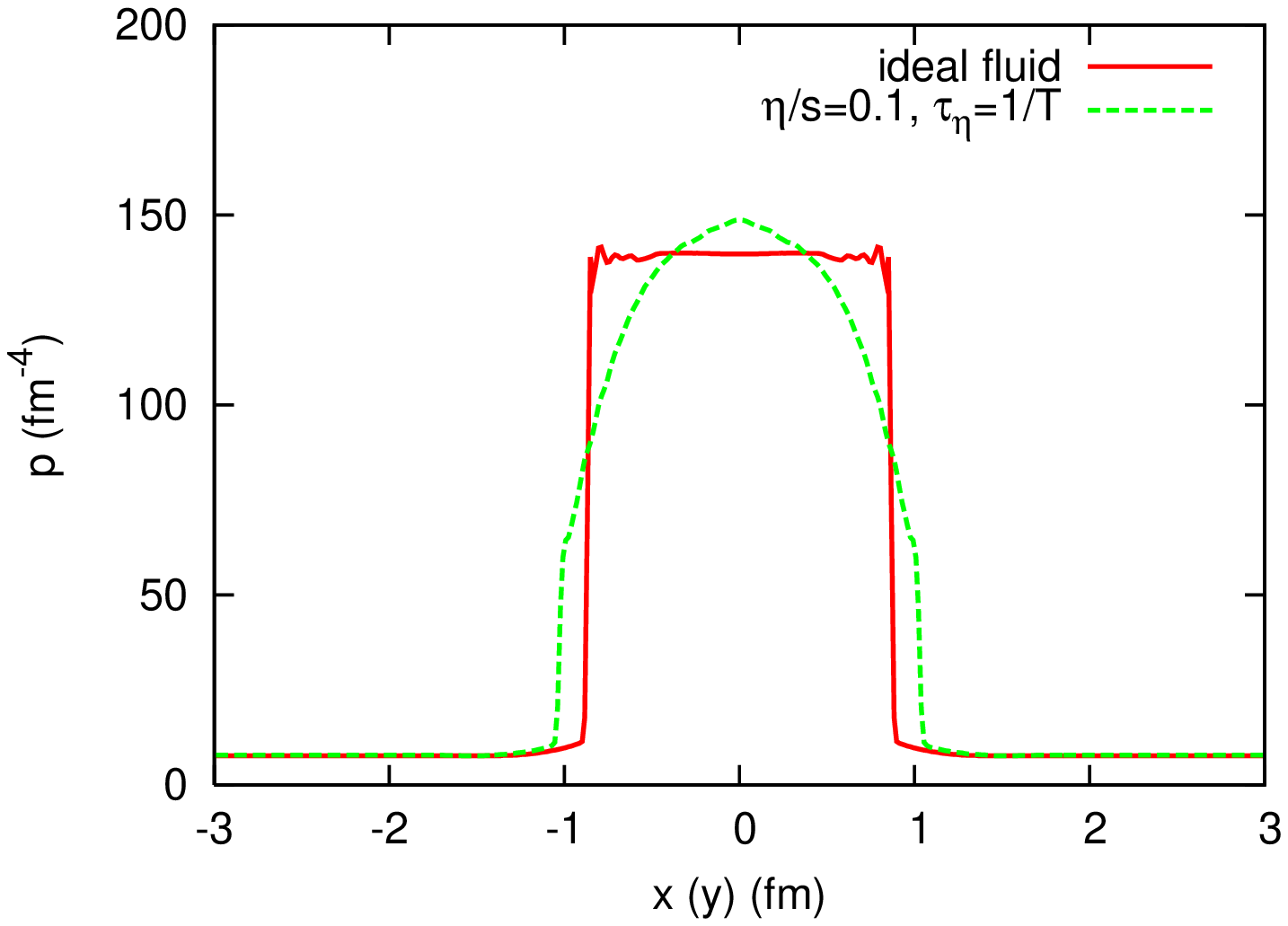}
\includegraphics[width=5.5cm,scale=0.9,angle=-90,clip]{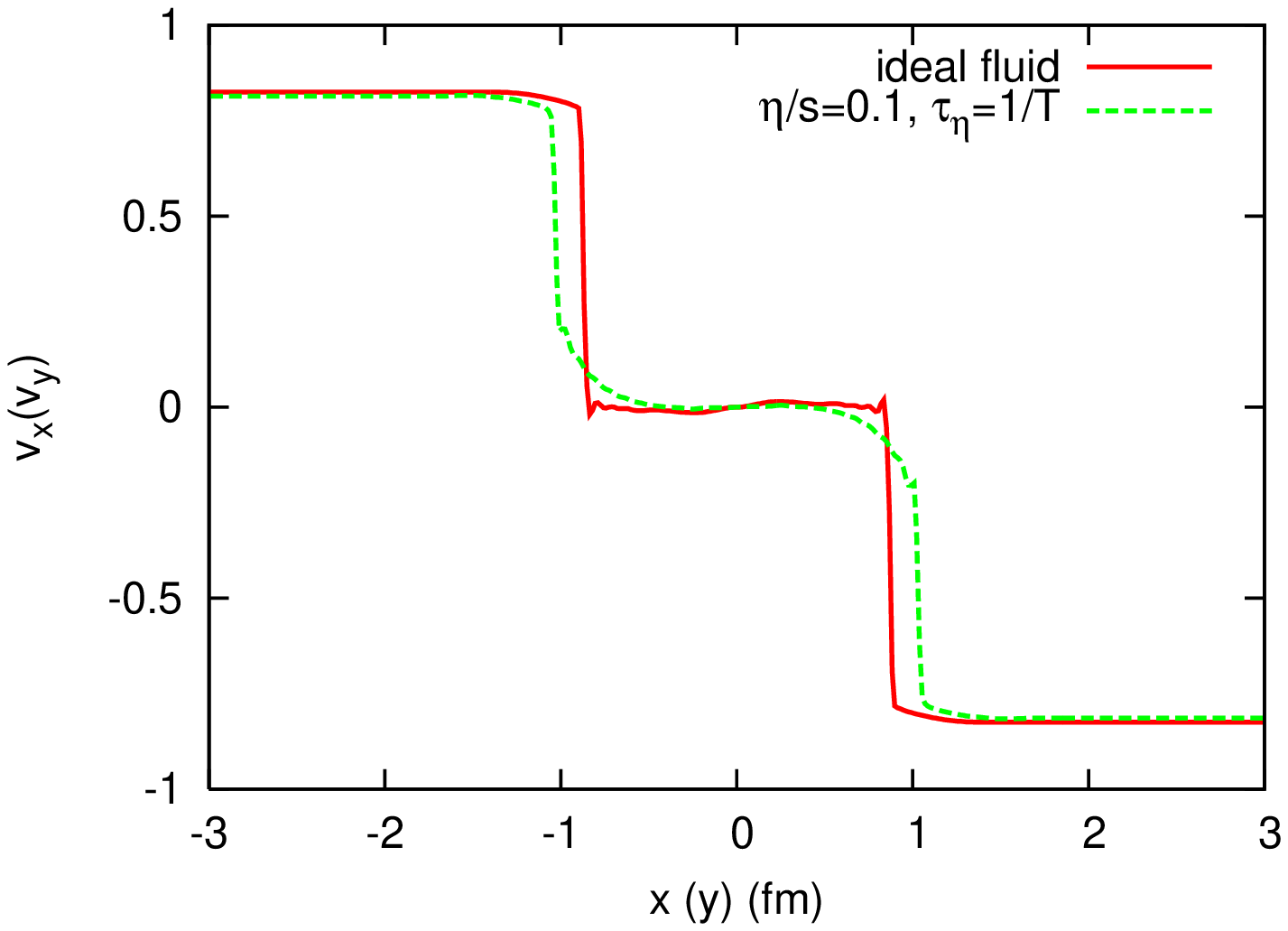}
\caption{
(Color online.)
Simulation of ideal and viscous hydrodynamics with lattice QCD EoS at $t=2.45$ fm (1500 steps).
The upper panels are two-dimensional profiles of (left) pressure and (right) flow velocity for the ideal hydrodynamic simulation.
The lower panels are one-dimensional profiles of (left) pressure and (right) $x(y)$-component of flow velocity at $y(x)=0$ fm for both ideal and viscous hydrodynamic simulations.
The finite viscous coefficient is $\eta/s=0.1$ and the relaxation time for the shear mode is $\tau_{\eta}=1/T$.
}
\label{fig:blast_qcd}
\end{figure}

In Fig.~\ref{fig:blast_qcd}, we show the results of simulation (ii) at $t=2.45$ fm (1500 steps).
In the upper panels we plot the pressure and velocity profiles for the ideal hydrodynamic simulation, and in the lower panels we show one-dimensional profiles of pressure and $x(y)$-component of flow velocity at $y(x)=0$ fm for both ideal and viscous hydrodynamic simulations.
Here we find qualitatively same features as in the simulation (i), but there are quantitative differences.
The pressure in the central region is about two times higher than that in the simulation (i).
The radius of the central region is about 10\% smaller than that in simulation (i).
The smaller radius is explained by the fact that the lattice QCD EoS is softer than the free gas EoS, as shown in Fig.~\ref{fig:eos}.
The pressure difference is explained by the ratio $e/p$ of the lattice QCD EoS at low and high temperatures as follows.
At low (high) temperature, this ratio is $e/p\sim 6 \ (e/p\sim 3)$, while it is $e/p=3$ for the free gas EoS.
Therefore, the energy density of the central region becomes about $6/3=2$ times larger than in simulation (i), and the pressure of this hot region is also about 2 times larger.

We have also successfully performed (3+1)-dimensional blast wave simulations for both ideal and viscous hydrodynamics with the same initial conditions $p_0=1 \ {\rm fm^{-4}}$, $v_r=0.9$ as in the (2+1)-dimensional simulations.
In the viscous hydrodynamic simulation, we choose the same parameterization for viscosity $\eta/s = 0.1$ and relaxation time $\tau_{\eta}=1/T$ as before.
Since these results were quite similar to those of the (2+1)-dimensional case, we do not show them here.

\section{Summary} 
\label{sec:sum} 
In this article, we have presented a state-of-the-art numerical algorithm for solving the relativistic viscous hydrodynamics equation with the QCD EoS. 
The numerical scheme is suitable for analyses of shock wave phenomena and has less numerical viscosity.
Both features are important for understanding feature of QGP features in high-energy heavy-ion collisions. 
We apply the algorithm to several numerical test problems, such as sound wave propagation, shock tube and blast wave problems. 
We investigated the precision of our numerical scheme in sound wave propagation using the free gas EoS and the lattice QCD EoS.
In both cases, the L1 norm scales as $\propto 1/N_{\rm cell}^2$ with the number of cells, which shows the second-order accuracy of our algorithm.
Moreover, we have estimated the numerical dissipation of our scheme $\eta_{\rm num}\approx [c_{\rm s}(e+p)/\lambda]\cdot(\Delta x)^2$, both in the presence and absence of physical viscosity.
We have shown the results of the shock tube test with our new numerical 
scheme, which suffers less numerical dissipation effect. This suggests that
this scheme is more suitable for analyses of physical viscosities
than SHASTA, which are currently mainly used in studies of high-energy heavy-ion collisions. 
We performed (2+1)- and (3+1)-dimensional blast wave simulations in ideal and viscous hydrodynamics with free gas EoS and lattice QCD EoS. 

The numerical scheme for relativistic viscous hydrodynamics with lattice QCD EoS is stable, capable of capturing the shock wave and has less artificial viscosity. 
These features create a solid baseline for comprehensive understanding of the QGP in high-energy heavy-ion collisions from the point of view of phenomenological analyses based on relativistic hydrodynamics. 

\section*{Acknowledgment}
We would like to thank Sangyong Jeon for the stimulating discussion during his stay at the Kobayashi-Maskawa Institute as a KMI visitor.
We would also like to thank Harri Niemi for providing us with their results 
shown in Fig.~\ref{fig:shock-tube-test}. 
This work was in part supported by
the Sasakawa Scientific Research Grant from the Japan Science Society,
Grant-in-Aid for Young Scientists (B) (22740156), 
Grant-in-Aid for Scientific Research (S)(22224003)
and the Kurata Memorial Hitachi 
Science and Technology Foundation.

\appendix
\section{Israel-Stewart formalism}
\label{sec:IS}
Let us summarize the Israel-Stewart formalism for causal viscous hydrodynamics.
In the relativistic viscous hydrodynamics in Landau-Lifshitz frame \cite{Landau:fluid}
\footnote{
Since we are interested in the quark-gluon plasma at vanishing chemical potential, the Eckart frame, in which the four-velocity $u^{\mu}$ is defined as $J_B^{\mu}=n_{\rm B} u^{\mu}$, is inconvenient.
This is because the baryon number current is $J_{\rm B}^{\mu}=(0,0,0,0)$ in equilibrium and hence $u^{\mu}$ becomes ill-defined.
},
the energy-momentum tensor $T^{\mu\nu}$ and the baryon number current $J_{\rm B}^{\mu}$ are decomposed as
\begin{eqnarray}
T^{\mu\nu}&=&eu^{\mu}u^{\nu}-(p+\Pi)\triangle^{\mu\nu}+\pi^{\mu\nu},\\
\label{eq:viscous_hydro}
J_{\rm B}^{\mu}&=& n_{\rm B} u^{\mu}+\nu_{\rm B}^{\mu},\\
\triangle^{\mu\nu}&\equiv& g^{\mu\nu}-u^{\mu}u^{\nu},
\end{eqnarray}
with bulk pressure $\Pi$, shear stress tensor $\pi^{\mu\nu}$, and dissipative baryon number current $\nu_{\rm B}^{\mu}$ which satisfy $\pi^{\mu\nu}u_{\nu}=0,\pi^{\mu}_{\mu}=0,\nu_{\rm B}^{\mu}u_{\mu}=0$.
Thermodynamic quantities $e,p$, and $n_{\rm B}$ are related through the equation of state $p=p(e,n_{\rm B})$ derived in equilibrium state and the terms that include $\Pi,\pi^{\mu\nu}$, and $\nu_{\rm B}^{\mu}$ make extra contributions to $T^{\mu\nu}$ and $J_{\rm B}^{\mu}$ in non-equilibrium state.
In the second-order formalism by Israel and Stewart \cite{Israel:1976tn}, the entropy current $s^{\mu}$ is constructed so that it is defined locally without derivatives, includes terms with dissipative quantities $(\Pi,\nu_{\rm B}^{\mu},\pi^{\mu\nu})$ up to second order, and must satisfy the condition that $s^{\mu}u_{\mu}$ is maximized in equilibrium $(\Pi,\nu_{\rm B}^{\mu},\pi^{\mu\nu}=0)$.
Such entropy current is then constructed by
\begin{eqnarray}
s^{\mu}=su^{\mu}-\frac{\mu_{\rm B}}{T}\nu_{\rm B}^{\mu}
-\frac{1}{T}(\alpha_0\Pi\nu_{\rm B}^{\mu}+\alpha_1\pi^{\mu\nu}\nu_{\rm B\nu})
-\frac{u^{\mu}}{2T}(\beta_0\Pi^2 -\beta_1\nu_{\rm B}^{\mu}\nu_{\rm B\mu}+\beta_2\pi^{\rho\sigma}\pi_{\rho\sigma}),
\end{eqnarray}
with coupling coefficients $\alpha_{0,1}$ and $\beta_{0,1,2} \ (\geq 0)$ and baryon number chemical potential $\mu_{\rm B}$.
In our algorithm, we neglect the couplings among different diffusive modes ($\alpha_0=\alpha_1=0$).
Calculating the divergence of the entropy current by using the conservation laws (\ref{eq-hydro}) and (\ref{eq-net_baryon}), we obtain
\begin{eqnarray}
\label{eq:entropy_production}
\partial_{\mu}s^{\mu}&=&
-\frac{\Pi}{T}(\triangle^{\mu\nu}\partial_{\mu}u_{\nu}+\alpha_0\partial_{\mu}\nu_{\rm B}^{\mu}+\beta_0\dot\Pi)
+\frac{\pi^{\mu\nu}}{T}(\partial_{\mu}u_{\nu}-\alpha_1\partial_{\mu}\nu_{\rm B\nu}-\beta_2\dot\pi_{\mu\nu})\nonumber\\
&&-\frac{\nu_{\rm B}^{\mu}}{T}\left[T\partial_{\mu}\left(\frac{\mu_{\rm B}}{T}\right)
+\alpha_0\partial_{\mu}\Pi+\alpha_1\partial_{\nu}\pi^{\nu}_{\mu}-\beta_1\dot\nu_{\rm B\mu}\right],
\end{eqnarray}
where $\dot f\equiv u^{\mu}\partial_{\mu}f$.
In deriving Eq.~(\ref{eq:entropy_production}), we neglect terms in higher-order deviation from equilibrium such as $-\frac{\partial_{\mu}u^{\mu}}{2T}\beta_0\Pi^2$.
In order to ensure that the entropy does not decrease, $\Pi$, $\pi_{\mu\nu}$, and $\nu_{\rm B}^{\mu}$ must obey the following constitutive equations:
\begin{eqnarray}
\label{eq:constitutive1}
-\Pi&=&\zeta(\triangle^{\mu\nu}\partial_{\mu}u_{\nu}+\alpha_0\partial_{\mu}\nu_{\rm B}^{\mu}+\beta_0\dot\Pi),\\
\label{eq:constitutive2}
\pi_{\mu\nu}&=&2\eta\langle\langle\partial_{\mu}u_{\nu}
-\alpha_1\partial_{\mu}\nu_{\rm B\nu}
-\beta_2\dot\pi_{\mu\nu}\rangle\rangle,\\
\label{eq:constitutive3}
\nu_{\rm B}^{\mu}&=&\sigma \triangle^{\mu\rho}\left[T\partial_{\rho}\left(\frac{\mu_{\rm B}}{T}\right)
+\alpha_0\partial_{\rho}\Pi+\alpha_1\partial_{\sigma}\pi^{\sigma}_{\rho}
-\beta_1\dot\nu_{\rho}\right],
\end{eqnarray}
with bulk and shear viscosities $\zeta, \ \eta \ (\geq0)$ and baryon number conductivity $\sigma \ (\geq 0)$.
Here $\langle\langle X^{\mu\nu}\rangle\rangle$ denotes a spatial, symmetric, and traceless tensor extracted from a general tensor $X^{\mu\nu}$:
\begin{eqnarray}
\langle\langle X^{\mu\nu}\rangle\rangle\
\equiv \triangle^{\mu\rho}\triangle^{\nu\sigma}
\left[\frac{X_{\rho\sigma}+X_{\sigma\rho}}{2}-\frac{\triangle_{\rho\sigma}\triangle^{\alpha\beta}X_{\alpha\beta}}{3}\right].
\end{eqnarray}
Note that the diffusive modes $\Pi,\pi^{\mu\nu}$, and $\nu_{\rm B}^{\mu}$ are now dynamical degrees of freedom and relax toward the first-order Navier-Stokes values:
\begin{eqnarray}
\Pi_{\rm NS}\equiv -\zeta\Delta^{\mu\nu}\partial_{\mu}u_{\nu}, \ \ 
\pi_{\rm NS}^{\mu\nu}\equiv 2\eta\langle\langle \partial^{\mu}u^{\nu} \rangle\rangle, \ \ 
\nu_{\rm B,NS}^{\mu}\equiv \sigma\Delta^{\mu\rho}T\partial_{\rho}\left(\frac{\mu_{\rm B}}{T}\right).
\end{eqnarray}
The relaxation times for these diffusive modes are given by
\begin{eqnarray}
\tau_{\zeta} \equiv \beta_0\zeta, \ \tau_{\eta} \equiv 2\beta_2\eta, \
\tau_{\sigma} \equiv  \beta_1\sigma.
\end{eqnarray}
Setting the couplings $\alpha_0=\alpha_1=0$ for simplicity, the Israel-Stewart equations are derived from the constitutive equations:
\begin{eqnarray}
\label{eq:Israel-Stewart1}
\left(\partial_t + \bm v\cdot \bm\partial\right) \Pi
&=& -\frac{\Pi-\Pi_{\rm NS}}{\gamma\tau_{\zeta}},\\
\label{eq:Israel-Stewart2}
\left(\partial_t + \bm v\cdot \bm\partial\right) \pi^{\mu\nu}
&=& -\frac{\pi^{\mu\nu}-\pi^{\mu\nu}_{\rm NS}}{\gamma\tau_{\eta}}
- \frac{1}{\gamma}\left(\pi^{\mu\alpha}u^{\nu}+\pi^{\nu\alpha}u^{\mu}\right)\dot u_{\alpha},\\
\label{eq:Israel-Stewart3}
\left(\partial_t + \bm v\cdot \bm\partial\right) \nu_{\rm B}^{\mu}
&=& -\frac{\nu_{\rm B}^{\mu}-\nu_{\rm B, NS}^{\mu}}{\gamma\tau_{\sigma}}
- \frac{1}{\gamma}\left(q^{\alpha}\dot u_{\alpha}\right)u^{\mu},
\end{eqnarray}
where the nonlinear terms in Eqs.~\eqref{eq:Israel-Stewart2} and \eqref{eq:Israel-Stewart3} (the second terms on the right hand sides) come from the constraints on $\pi^{\mu\nu}$ and $\nu_{\rm B}^{\mu}$.
Although Eqs.~\eqref{eq:Israel-Stewart2} and \eqref{eq:Israel-Stewart3} are equivalent to Eqs.~\eqref{eq:constitutive2} and \eqref{eq:constitutive3} ($\alpha_0=\alpha_1=0$), the nonlinear terms in the formers do not contain all of the higher-order terms of the same order and so we consider only the linear part for spatial components:
\begin{eqnarray}
\label{eq:Israel-Stewart-lin2}
\left(\partial_t + \bm v\cdot \bm\partial\right) \pi^{ij}
&=& -\frac{\pi^{ij}-\pi^{ij}_{\rm NS}}{\gamma\tau_{\eta}},\\
\label{eq:Israel-Stewart-lin3}
\left(\partial_t + \bm v\cdot \bm\partial\right) \nu_{\rm B}^{i}
&=& -\frac{\nu_{\rm B}^{i}-\nu_{\rm B, NS}^{i}}{\gamma\tau_{\sigma}}.
\end{eqnarray}
This approximation should work when the gradient of the fluid variables is not so steep.
Together with the above Israel-Stewart equations, the system evolves according to the conservation laws:
\begin{eqnarray}
\label{eq:conservation_vis}
\frac{\partial}{\partial t}
\left(\begin{array}{c}
D+\nu_{\rm B}^0\\
m^i-\Pi\Delta^{0i}+\pi^{0i}\\
E-\Pi\Delta^{00}+\pi^{00}
\end{array}\right)
+{\nabla_j}\cdot
\left(\begin{array}{c}
Dv^j + \nu_{\rm B}^j\\
m^i v^j + p\delta^{ij} - \Pi\Delta^{ij}+\pi^{ij}\\
m^j - \Pi\Delta^{0j}+\pi^{0j}
\end{array}\right)
=0.
\end{eqnarray}

\section{Numerical algorithm}
\label{sec:numerical_algorithm}
Here we will give a brief summary of our numerical algorithm of the causal viscous hydrodynamics based on Ref.~\cite{Takamoto:2011wi}.
Let us first introduce the conserved variables $\bm U=\bm U_{\rm id}+\bm U_{\rm vis}$, where $\bm U_{\rm id}=(D,\bm m, E)$ and $\bm U_{\rm vis}=(\nu_{\rm B}^0,\Pi^{0i},\Pi^{00})$ denote the contribution from the ideal and viscous components respectively.
In the causal viscous hydrodynamics, we use the primitive variables $\bm V_{\rm id} = (n_{\rm B},\bm v, p)$ for the ideal component and $\bm V_{\rm vis} = (\Pi^{ij},\nu^k_{\rm B}) \ (i\geq j, \ i,j,k=1,2,3)$ for the viscous component.
Here we define $\Pi^{\mu\nu}$ as the total viscous component in the energy-momentum tensor $\Pi^{\mu\nu}\equiv-\Pi\Delta^{\mu\nu}+\pi^{\mu\nu}$, which is transverse $\Pi^{\mu\nu}u_{\nu}=0$ and symmetric $\Pi^{\mu\nu}=\Pi^{\nu\mu}$.
Recall the relations among these variables $\bm U_{\rm id}=\bm U_{\rm id}(\bm V_{\rm id})$ and $\bm U_{\rm vis}=\bm U_{\rm vis}(\bm V_{\rm id}, \bm V_{\rm vis})$.
The time evolution by the causal viscous hydrodynamics is summarized by
\begin{eqnarray}
\label{eq:conservation}
\bm U(t) &\rightarrow& \bm U(t+\Delta t),\\
\label{eq:Israel-Stewart}
\bm V_{\rm vis}(t)&\rightarrow& \bm V_{\rm vis}(t+\Delta t),
\end{eqnarray}
where Eq.~\eqref{eq:conservation} represents the time evolution by the energy-momentum conservation laws and continuity equation for the baryon number (Eq.~\eqref{eq:conservation_vis}) and Eq.~\eqref{eq:Israel-Stewart} represents the evolution by the Israel-Stewart equation (Eqs.~\eqref{eq:Israel-Stewart1}, \eqref{eq:Israel-Stewart-lin2}, and \eqref{eq:Israel-Stewart-lin3}).
In the numerical algorithm, we evolve the primitive variables $\bm V_{\rm id, vis}$ according to these hydrodynamics equations.
Note that $\bm U$ and $\bm V_{\rm vis}$ carry enough information to obtain $\bm U_{\rm id}$, $\bm U_{\rm vis}$, and $\bm V_{\rm id}$ in principle.

Following Ref.~\cite{Takamoto:2011wi}, we construct a numerical algorithm for the causal viscous hydrodynamics.
We split the time evolution into 3 steps: 
\begin{enumerate}
\item
Ideal part of the conservation laws:
$\bm U_{\rm id}(t)\rightarrow \bm U_{\rm id}^*(t+\Delta t)$, $\bm V_{\rm id}(t)\rightarrow \bm V_{\rm id}^*(t+\Delta t)$.
\begin{eqnarray}
\frac{\partial}{\partial t}
\left(\begin{array}{c}
D\\
m^i\\
E
\end{array}\right)
+{\nabla_j}\cdot
\left(\begin{array}{c}
Dv^j\\
m^i v^j + p\delta^{ij}\\
m^j
\end{array}\right)
=0.
\end{eqnarray}
Here we evolve only $\bm U_{\rm id}(t)$ by using the currents of the ideal component (Eq.~\eqref{eq:conservation_laws}).
The Riemann solver we proposed is used in evaluating the numerical flux to obtain the currents.
In the ideal hydrodynamics, we only need this step.
In the viscous hydrodynamics, this step does not give $\bm U_{\rm id}$ and $\bm V_{\rm id}$ at time $t+\Delta t$ as indicated by asterisks $^*$.
Note that in this step the total conserved quantities vary from $\bm U(t)=\bm U_{\rm id}(t)+\bm U_{\rm vis}(t)$ to $\bm U_{\rm id}^*(t+\Delta t)+\bm U_{\rm vis}(t)$ so that this step satisfies the conservation law.
In multi-dimensional case, we evolve by the dimensional splitting method.

\item
Israel-Stewart equation:
$\bm V_{\rm vis}(t)\rightarrow \bm V_{\rm vis}(t+\Delta t/2)\rightarrow \bm V_{\rm vis}(t+\Delta t)$.
\begin{eqnarray}
\label{eq:split1}
\left(\partial_t + \bm v\cdot \bm\partial\right) \Pi
= 0, \ \
\left(\partial_t + \bm v\cdot \bm\partial\right) \pi^{ij}
= 0, \ \ 
\left(\partial_t + \bm v\cdot \bm\partial\right) \nu_{\rm B}^{i}
= 0,
\end{eqnarray}
\begin{eqnarray}
\label{eq:split2}
\partial_t \Pi
= -\frac{\Pi-\Pi_{\rm NS}}{\gamma\tau_{\zeta}}, \ \ 
\partial_t \pi^{ij}
= -\frac{\pi^{ij}-\pi^{ij}_{\rm NS}}{\gamma\tau_{\eta}}, \ \
\partial_t \nu_{\rm B}^{i}
= -\frac{\nu_{\rm B}^{i}-\nu_{\rm B, NS}^{i}}{\gamma\tau_{\sigma}}.
\end{eqnarray}
This step consists of solving the advection part (Eq.~\eqref{eq:split1}) and the relaxation part (Eq.~\eqref{eq:split2}) of the Israel-Stewart equation separately and implementing them by the operator splitting method as in Ref.~\cite{Takamoto:2011wi}.
In the relaxation part, we calculate the time differentiation of $\bm V_{\rm id}$ in the Navier-Stokes terms by $\left[\bm V_{\rm id}^*(t+\Delta t)-\bm V_{\rm id}(t)\right]/\Delta t$.
We use $\bm V_{\rm id}^*(t+\Delta t)$ to evaluate $\bm V_{\rm id}$ in all the other parts in the Israel-Stewart equation.
We use the intermediate state $\bm V_{\rm vis}(t+\Delta t/2)$ in the next step.
In multi-dimensional case, the advection part is solved by the dimensional splitting method while the relaxation time cannot be dimensionally split because of the Navier-Stokes terms.

\item
Viscous part of the conservation laws:
$\bm U_{\rm id}^*(t+\Delta t) + \bm U_{\rm vis}(t)\rightarrow \bm U(t+\Delta t)$.
\begin{eqnarray}
\frac{\partial}{\partial t}
\left(\begin{array}{c}
D+\nu_{\rm B}^0\\
m^i-\Pi\Delta^{0i}+\pi^{0i}\\
E-\Pi\Delta^{00}+\pi^{00}
\end{array}\right)
+{\nabla_j}\cdot
\left(\begin{array}{c}
\nu_{\rm B}^j\\
- \Pi\Delta^{ij}+\pi^{ij}\\
- \Pi\Delta^{0j}+\pi^{0j}
\end{array}\right)
=0.
\end{eqnarray}
Here we evolve the sum $\bm U_{\rm id}^*(t+\Delta t) + \bm U_{\rm vis}(t)$ by  the currents of the viscous component using $\bm V_{\rm vis}(t+\Delta t/2)$ and $\bm V_{\rm id}^*(t+\Delta t)$ to obtain $\bm U(t+\Delta t)$.
From $\bm U(t+\Delta t)$ and $\bm V_{\rm vis}(t+\Delta t)$, we get $\bm U_{\rm id}(t+\Delta t)\neq \bm U_{\rm id}^*(t+\Delta t)$ and $\bm V_{\rm id}(t+\Delta t)\neq \bm V_{\rm id}^*(t+\Delta t)$ in general.
For details of primitive recovery for viscous hydrodynamics, see Ref.~\cite{Takamoto:2011wi}.
Note that this step also satisfies the conservation law.
In multi-dimensional case, we evolve by the dimensional splitting method.
\end{enumerate}

The $\Delta t$ is defined so that it satisfies the CFL condition of the telegrapher equation as in Ref.~\cite{Takamoto:2011wi}.
This numerical algorithm is applicable to both Landau-Lifshitz and Eckart frames of causal viscous hydrodynamics.

In our algorithm, we approximate the spatial derivatives of the Navier-Stokes terms with the centered finite differences because the physical meanings of the dissipation variables are the diffusion.
For the other part of the spatial derivatives, we utilize the MUSCL scheme by Van Leer \cite{VanLeer} for the second-order accuracy in space.
By this numerical algorithm, we achieve the second-order accuracy in both space and time as we checked in Section \ref{sec:num_test}.

\section{Comparison of Landau-Lifshitz and Eckart frames}
\label{sec:landau_eckart}

\begin{figure}[t!]
\includegraphics[width=5cm,scale=0.5,angle=-90,clip]{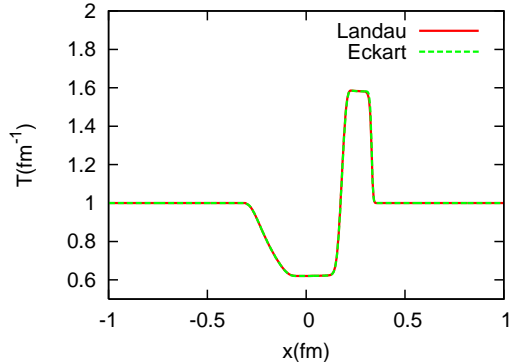}
\caption{
(Color online.)
Temperature of the shock tube problem at $t=2.7$ fm (1400 steps) in Landau-Lifshitz and Eckart frames.
The particle diffusion in Landau-Lifshitz frame, or the heat conductivity in Eckart frame, is $\kappa=0.05 \ {\rm fm}^{-2}$.
}
\label{fig:EL_test}
\end{figure}

Here we compare simulations of viscous hydrodynamics in Landau-Lifshitz and Eckart frames.
The purpose of this comparison is to show that the original hydrodynamic code in Eckart frame \cite{Takamoto:2011wi} is extended correctly to a code in Landau-Lifshitz frame.
Therefore we perform the simulation with the equation of state with high baryon density as in \cite{Takamoto:2011wi} and we do {\it not} use the Riemann solver that we propose in the text.

We simulate a shock tube problem with an initial condition:
\begin{eqnarray}
\bm{V}(x,t=0)
=\Biggl\{\begin{array}{lc}
\bm V_L = (d_{L0},0,v_{y0},0,p_{L0}) & (x<0) \\
\bm V_R = (d_{R0},0,-v_{y0},0,p_{R0}) & (x>0)
\end{array},
\end{eqnarray}
with $d_{L0}=p_{L0}=10 \ {\rm fm^{-4}}$, $d_{R0}=p_{R0}=1 \ {\rm fm^{-4}}$, and $v_{y0}=0.2$.
Note that $d(x)$ denotes the mass density in this simulation.
The system length is 2 fm and is discretized with 200 cells.
The particle diffusion in Landau-Lifshitz frame, or equivalently the heat conductivity in Eckart frame, is $\kappa=0.05 \ {\rm fm}^{-2}$ and shear and bulk viscous coefficients are $\eta=\zeta=0 \ {\rm fm}^{-3}$.

Shown in Fig.~\ref{fig:EL_test} is the temperature at $t=2.7$ fm (1400 steps) in Landau-Lifshitz and Eckart frames.
According to \cite{Israel:1976tn}, the difference in thermodynamic quantities in these frames is small.
Since we cannot find frame dependence in the temperature profiles, we conclude that we have successfully extended the original code to the one in Landau-Lifshitz frame.

\vspace{-0.4cm}
\bibliographystyle{apsrev}

\end{document}